\pdfoutput=1

\documentclass[12pt]{amsart}

\usepackage{pdfsync}
\usepackage{subfigure}

\usepackage{latexsym,amsmath,amsthm,amsfonts, bbm}
\usepackage[psamsfonts]{amssymb}
\usepackage{amssymb,epsf,lscape,color,colortbl,subfigure,epsfig, setspace}
\usepackage{amsmath, amsthm, amssymb, color, fancybox, epsfig, subfigure, float,epsf,lscape,colortbl}

\newtheorem{conj}{Conjecture}
\newtheorem{thm}[conj]{\sc Theorem}

\newtheorem{prop}[conj]{\sc Proposition}
\newtheorem{lemma}[conj]{Lemma}

\def\to{\rightarrow}
\def\II{\mathbb I}
\def\EE{ {\rm I} \kern-.15em {\rm E} }
\def\PP{ {\rm I} \kern-.15em {\rm P} }

\def\wh{\widehat}
\def\eps{\varepsilon}

\def\T{\mbox{$\mathcal T$}}

\def\RR{\mathbb R}
\def\EE{\mathbb{E}}
\def\PP{\mathbb{P}}
\def\1{\mathbbm{1}}

\def\eps{\varepsilon}

\begin{document}
\title{Adaptive inference for the mean of a stochastic process in functional data}

\author[Bunea, Wegkamp and Ivanescu]{ Florentina Bunea$^1$, Marten H. Wegkamp$^1$, Andrada E. Ivanescu$^2$}

\footnotetext[1]{
Department of Statistics, Florida State University, Tallahassee, FL 32306-4330. The research of Florentina Bunea and Marten
  Wegkamp was supported in part by NSF Grant DMS-0706829. Part of the research was done while the authors were visiting the Isaac Newton Institute for Mathematical Sciences  (Statistical Theory and Methods for Complex, High-Dimensional Data Programme) at Cambridge University during Spring 2008.}
\footnotetext[2]{Department of Biostatistics, East Carolina University, Greenville, NC 27858-4353.}

\maketitle

\begin{abstract}
This paper  proposes and analyzes fully data driven methods for inference about the mean function of a stochastic process from  a sample  of independent trajectories of the process, observed at discrete time points and corrupted by additive random error.  
The proposed method uses thresholded least squares estimators relative to an approximating function basis. The variable threshold levels are estimated from the data and the basis is chosen via cross-validation from a library of bases.  The resulting estimates adapt to the unknown sparsity of the mean function relative to the selected approximating basis,  both in terms of the mean squared error and supremum norm. These results are based on novel oracle inequalities. In addition, uniform confidence bands for the mean function of the process are constructed.  The bands also adapt to the unknown regularity of the mean function, are  easy to compute, and do not require explicit estimation of the covariance operator of the process.  The simulation study that complements the theoretical results shows that the new method performs very well in practice, and is robust against large variations introduced by the random error terms.\\

\end{abstract}

\baselineskip=18pt
\section{Introduction}

In this paper we develop and analyze new methodology for inference about  the mean of a stochastic process from data that consists of  
independent realizations  of  a stochastic  process observed at discrete times,   where each observation   is contaminated by an additive
error term.  Formally, let  $\{ X(t), \ 0\le t\le 1\}$ be a stochastic process  with mean function $$f(t)= \EE [X(t)]$$
and covariance function $$\Gamma(s,t)= \text{Cov}(X(s), X(t)),$$ for all $0\le s, t\le 1$.  We denote the zero mean process $X(t)-f(t)$ by $Z(t)$.
We observe $Y_{ij}$  at times $t_{j}$, for $1\le i\le n$, $1\le j\le m$, that are of the form
\begin{eqnarray}\label{model}
Y_{ij} &=& X_i(t_{j}) + \eps_{ij} 
\end{eqnarray}
where $X_i(t)$, with mean $f(t)$,  are random  independent realizations of the process $X(t)$. We assume that 
$\eps_{ij}$ are independent across $i$ and $j$ with zero mean and variance
$\EE [\eps_{ij}^2] = \sigma_{\eps}^2$.  \\

In this paper we propose new methods for estimating and constructing confidence bands for $f$. Although the estimation of $f$ received considerable attention over the last decade, the theoretical study of data adaptive estimators in model (\ref{model})   is still open to investigation.  In contrast with the abundance of methods for estimating $f$, methods  for constructing confidence bands for $f$ are very limited. This motivates  our  twofold contribution to  the existing literature: (1) We construct computationally efficient  and fully data-driven estimators and confidence bands for  $f$, without making distributional assumptions on the process $Z(t)$ or smoothness assumptions on $f$; (2) We  assess the quality of our data adaptive estimates theoretically and prove that both the estimators and the confidence bands adapt to the unknown regularity of  $f$. Moreover, we show that our bands are,  asymptotically in $n$,  uniform  in $f$. \\

In what follows we review the existing results in the literature and provide further motivation for our procedure. The problem of estimating $f$ from data generated from (\ref{model}) has been considered by a large number of authors, starting with Ramsay and Silverman (2002, 2005) and Rupert, Wand and Carroll  (2003). The existing methods are either based on kernel smoothers [see, e.g., Zhang and Chen (2007), Yao (2007), Benko, H\"ardle and Kneip (2009)],  penalized splines [see,  e.g., Ramsay and Silverman (2005)], free-knot splines [see, e.g., Gervini (2006)], or ridge-type least squares estimates,  [see, e.g., Rice and Silverman (1991)].  All  resulting estimates  depend on tuning parameters that are method specific.  Theoretical properties of these estimates of $f$ are still emerging, and  have only been established for non-adaptive choices of the respective tuning parameters, that is choices that require prior knowledge of the smoothness of $f$,  [see, e.g. Zhang and Chen (2007) and Gervini (2006)].  Although guidelines for 
data-driven choices of these  parameters are offered in all these works, the
theoretical  properties of the resulting estimates are still open to investigation. 
In contrast, we suggest in Section 2 below  a computationally simple method based on thresholded least squares estimators. Our method  does not require any specification of the regularity of $f(t)$ or $X(t)$  prior to estimation. We show via oracle inequalities that  our estimators adapt to this unknown regularity.\\

 Whereas the estimation of the mean $f(t)$ of the process $X(t)$ is well understood, modulo the technical and possibly computational issues raised above, the construction of uniform confidence intervals for $f$ has not been investigated  in this context and in general the construction 
 of confidence bands for $f$ in model (\ref{model}) seems to have received little attention.  Zhang and Chen (2007) and Yao (2007) construct kernel-type estimators and show that they are asymptotically normal with mean $f(t)$ and variance $ \Gamma(t, t)/n$.

 Although not addressed  implicitly in these works, one can use these results to build confidence bands. This construction  would require the estimation of $\Gamma(t, t)$, using for instance the large  body of work on the estimation of the covariance operator, based  on the Karhunen-Loeve decomposition of the process $X(t)$ and the subsequent estimation the the functional eigenfunctions and eigenvalues, see, e.g., M\"uller (2005),  Benko, H\"ardle and Kneip (2009), who also comment on the possible instability of these estimates and offer refined methods for improved performance. We offer an alternative  method in Section 2.5 below. Our procedure is  computationally simple,  avoids direct estimation of the covariance matrix $\Gamma(t_j,t_k)$, $1\le j,k\le m$, and leads to adaptive bands that are uniform in the parameter $f$. \\
 

The rest of the paper is organized as follows. In Section 2.2 below we discuss thresholded least squares estimators in the  functional data setting. Our emphasis is on hard threshold estimators, but we also  discuss briefly the closely related  soft threshold estimators. In Section 2.3 we establish 
oracle inequalities for the fit of the estimators which show that the estimates adapt to the unknown sparsity of the mean $f$. The sparsity of $f$ is relative to a given approximating basis. In Section  2.4 we suggest cross-validation for choosing  the basis  from a library of bases. Since each basis induces an estimator, our procedure can be regarded as one of selecting an estimator from a given list. We then establish an oracle inequality for the selected estimator, that  shows that the selected estimator performs, essentially, as well as the best estimator from the list, in terms of mean squared error relative to the unknown $f$. In Section 2.5 we give the construction of the confidence bands and prove that they have the desired coverage probability.  Section 3 contains a comprehensive simulation study that  strongly supports  the theoretical merits of the method and indicates that our method compares favorably with existing methods. The net merit of the proposed  method is very visible 
when the variance of the random noise $\varepsilon$  is at the same level as that of the stochastic process
 $Z(t)$ and we discuss this in detail in Sections 3.2 and 3.3. 
All the proofs are collected in the Appendix.
\medskip

\section{Methodology}

\subsection{Preliminaries}
As explained in the introduction, the aim of this paper is (a) to estimate the mean $f(t)$ of the process $X(t)$ and (b) to construct confidence bands for the mean $f(t)$. Our approach is based on thresholded least squares estimates obtained relative to  bases $\phi_1,\ldots,\phi_m$ that are orthonormal  in  $L^2(\PP_m)$, where $\PP_m$ is the empirical measure that  puts mass $1/m$ at each $t_j$. Thus, our bases satisfy 
\begin{equation*}\label{ort} \frac1m \sum_{j=1}^m \phi_k(t_j) \phi_{k'}(t_j)= \1{\{k=k'\}}\end{equation*}
for $1\le k,k'\le m$. Examples include  the Fourier, local trigonometric and Haar bases.\\

Since $\phi_1,\ldots,\phi_m$ is orthonormal in $L^2(\PP_m)$, each $X_i(t_j) \equiv f(t_j) + Z_i(t_j)$ has the decomposition
\begin{eqnarray}\label{Xi}
 X_i(t_j)  &=&  \sum_{k=1}^{m}\mu_{k}\phi_k(t_j)  + \sum_{k=1}^{m}A_{ik}\phi_k(t_j),\ 1 \leq j \leq m,
\end{eqnarray}
with
\begin{equation}\label{mu}  \mu_k = \frac{1}{m}\sum_{j=1}^{m}f(t_{j})\phi_k(t_j), \end{equation}
and  
\begin{equation}\label{Aik}A_{ik} = \frac{1}{m}\sum_{j=1}^{m} Z_i(t_j)\phi_k(t_j).\end{equation}
For ease of notation we suppress the dependence on $m$ in $\mu_k$ and $A_{ik}$. 
Under model (\ref{model})  the random variables $A_{1k},\ldots, A_{nk}$, for each $k$, are independent and identically distributed with mean zero and variance  
\begin{eqnarray}\label{var}
\sigma_k^2 &\equiv& \EE[ A_{ik}^2 ] = \frac{1}{m^2} \sum_{j=1}^m \sum_{j'=1}^m \Gamma (t_j, t_{j'}) \phi_k(t_j) \phi_k(t_{j'}); 
\end{eqnarray}
in the special case where $\Gamma(s,t)=\tau^2 \1\{s=t\}$ for all $s,t$, the variances $\sigma_k^2$ reduce to $\sigma_k^2=\tau^2/m$  for $k=1,\ldots,m$.
The coefficients $\mu_k$ determine the target  vector $(f(t_1),\ldots, f(t_m))^\prime$
via the formula 
\begin{eqnarray}
f(t_j) =   \sum_{k=1}^{m}\mu_{k}\phi_k(t_j), \end{eqnarray}
 for each $j$. 
 We motivate  below our proposed methods for inference on  $(f(t_1),\ldots, f(t_m))^\prime$.\\

\subsection{Threshold-type estimators for functional data}

\noindent 
Our procedure falls between two of the currently used strategies: averaging estimated individual trajectories and applying various smoothing methods to the entire data set. Our initial estimator of  $f(t_j)$ is  a least squares estimator,  which  can 
 be viewed as an average (over $n$) of weighted values of the $Y_{ij}$'s. Our final estimator
will be a truncated version of the least squares estimator, with data dependent truncation levels determined from the entire data set. We describe our procedure below. The least squares estimator based on all observations of $\mu = (\mu_1, \ldots, \mu_k, \ldots \mu_m)$, for  $\mu_k$ defined by (\ref{mu}), is the vector $\wh\mu=(\wh\mu_1,\cdots,\wh\mu_m)$ that minimizes
 \[ \frac1n \sum_{i=1}^n\frac1m \sum_{j=1}^m \left\{ Y_{ij}-\sum_{k=1}^m \mu_k
\phi_k(t_j) \right\}^2\]  over $\mu \in\RR^m$. Using the orthonormality property of the 
basis the estimators $\wh\mu_k$ of $\mu_k$  are given by 
\begin{equation}\label{muhat} \wh{\mu}_k=\frac1n \sum_{i=1}^n \wh\mu_{i,k}, \end{equation}
 the sample average (over $n$) of $\wh\mu_{i,k}$, which  in turn are  the least squares estimators of $\mu_{i, k} = \mu_k + A_{ik}$ 
based on the observations $Y_{ij}$ from the $i$-th curve only. Using again the orthonormality property of the basis used for this fit, the estimators $\wh \mu_{i,k}$  of $\mu_{i,k}$ are given by 
\begin{equation}\label{part} \wh\mu_{i,k}=\frac1m \sum_{j=1}^m Y_{ij}\phi_k(t_j).
\end{equation}
Recalling that each $Y_{ij}$ follows model  (\ref{model}), and using the representations (\ref{Xi}) -- (\ref{Aik}), we can further write
$\wh\mu_{i,k}$ as \begin{eqnarray*} \wh\mu_{i,k}  &=&
\mu_k + A_{ik}+ \frac1m\sum_{j=1}^m \eps_{ij} \phi_k(t_j).
\end{eqnarray*}
Since $\eps_{ij}$ and $A_{ik}$ have mean zero and are independent across $i$ and $j$ we obtain, for every $k$,  that 
 \[ \EE[\wh\mu_{i,k}]=\mu_k \ \ \mbox{and}  \ \
\text{Var}(\wh\mu_{i,k})=\sigma_k^2 + \frac{\sigma_\eps^2 }{m}.\]
Similarly, we find that
 \begin{eqnarray*} \wh\mu_{k}  &=&
\mu_k +\frac1n\sum_{i=1}^n A_{ik}+ \frac1n
\sum_{i=1}^n\frac1m\sum_{j=1}^m \eps_{ij} \phi_k(t_j)
\end{eqnarray*}
with \[ \EE[\wh\mu_{k}]=\mu_k \ \  \mbox{and}  \ \ 
\text{Var}(\wh\mu_{k})=\frac{\sigma_k^2}{n} + \frac{\sigma_\eps^2
}{mn}.\] 
The initial (unbiased) estimator $\wh f_{LS}(t_j)$ based on the least squares estimates $\wh\mu_k$ of the mean function $f(t_j)$ is simply
\[ \wh f_{LS}(t_j)= \sum_{k=1}^m \wh \mu_k \phi_k(t_j),\ 1\le j\le m,
\]
\noindent and its variance may be  unnecessarily  inflated  by the presence of, possibly many, very small estimates $\wh \mu_k$. This can be remedied by truncating the coefficients at a level that takes into account both the variability of the measurement errors $\eps$ and the variability of the stochastic process $Z(t)$; this is the  essential difference between truncated estimators based on data generated as in (\ref{model}) and their counterpart based only on independent data in a standard nonparametric regression setting. We will use the truncation level $\wh r_k$ given below and we  will justify it theoretically and practically in the next sections.  Let $z(\alpha)$ be the quantile corresponding to a  $\mathcal{N}(0,1)$ random variable and $0 < \alpha < 1$ is a given number. Define 
\begin{equation}\label{trunc} \wh r_k=  \frac{S_k + \delta}{\sqrt{n}}z(\frac{\alpha}{2m}) ,\end{equation}
for some small $\delta>0$ (that is set to zero in practice) that depends on   
\begin{eqnarray*}
S_k^2 &=& \frac{1}{n-1} \sum_{i=1}^n ( \wh\mu_{i,k}- \wh\mu_k )^2.
\end{eqnarray*}
\noindent Notice that $S_k^2$ a consistent and unbiased estimator of  $\text{Var}(\wh\mu_{ik})= \sigma_k^2+
\sigma_\eps^2/m$ since the $\wh \mu_{i,k}$, $1\le i\le n$ are i.i.d. for fixed $1\le k\le m$. \\

We will focus on  hard threshold estimators of the coefficients $\mu_k$ and function $f$. They are, respectively 
\[ \wh\mu_k(\wh r_k) =:  \wh\mu_k \1\{ |\wh \mu_k |\ge \wh r_k\};  \ \ \   \wh f_{(\wh r)}=:  \sum_{k=1}^m \wh \mu_k(\wh r_k) \phi_k.\]

\noindent We will also consider, for completeness, the soft threshold estimators:  
\[\widetilde \mu_k(\wh r_k) =:  \text{sgn}(\wh\mu_k) ( |\mu_k|- \wh r_k)_{+}; \ \ \  \widetilde f_{(\wh r)}=:  \sum_{k=1}^m \widetilde \mu_k(\wh r_k) \phi_k .\]
The two estimators are closely related as the coefficients $\wh\mu_k(\wh r_k)$ and $\widetilde \mu_k(\wh r_k)$ differ by at most $\wh r_k$ since  $$ \widetilde \mu_k(\wh r_k) = \wh\mu_k(\wh r_k) - \text{sgn}(\wh\mu_k)(\wh r_k). $$

\noindent In the next section we discuss the goodness-of-fit of these estimates in terms of the mean squared error and  error in the supremum norm. 

\subsection{Oracle inequalities for the estimators of the mean of a stochastic process}

\begin{figure}[]
  \begin{center}
    \subfigure[Fourier basis approximation]{\label{fig:stoch-a}\includegraphics[scale=0.3]{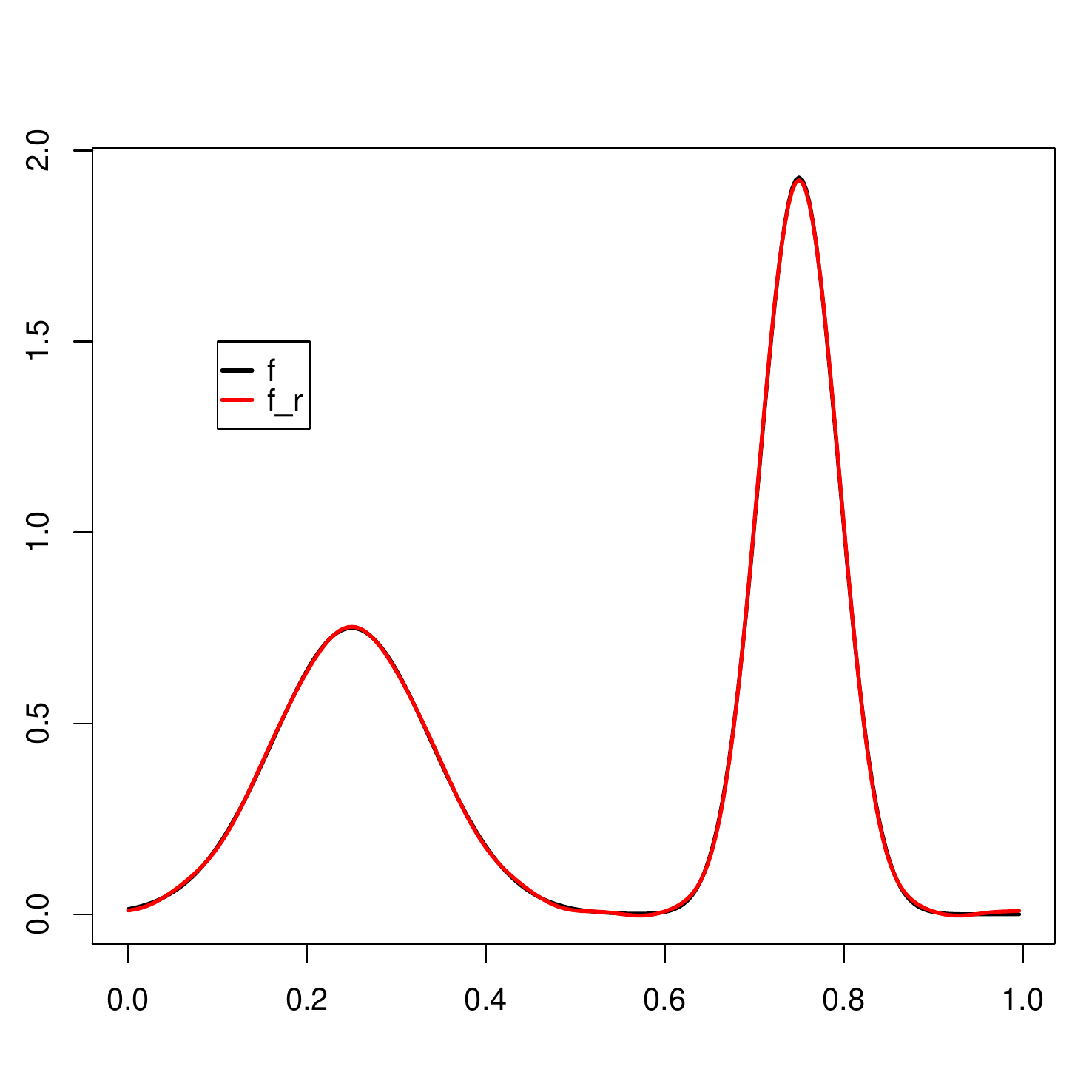}}
    \subfigure[Corresponding nonzero coeffcients]{\label{fig:stoch-b}\includegraphics[scale=0.3]{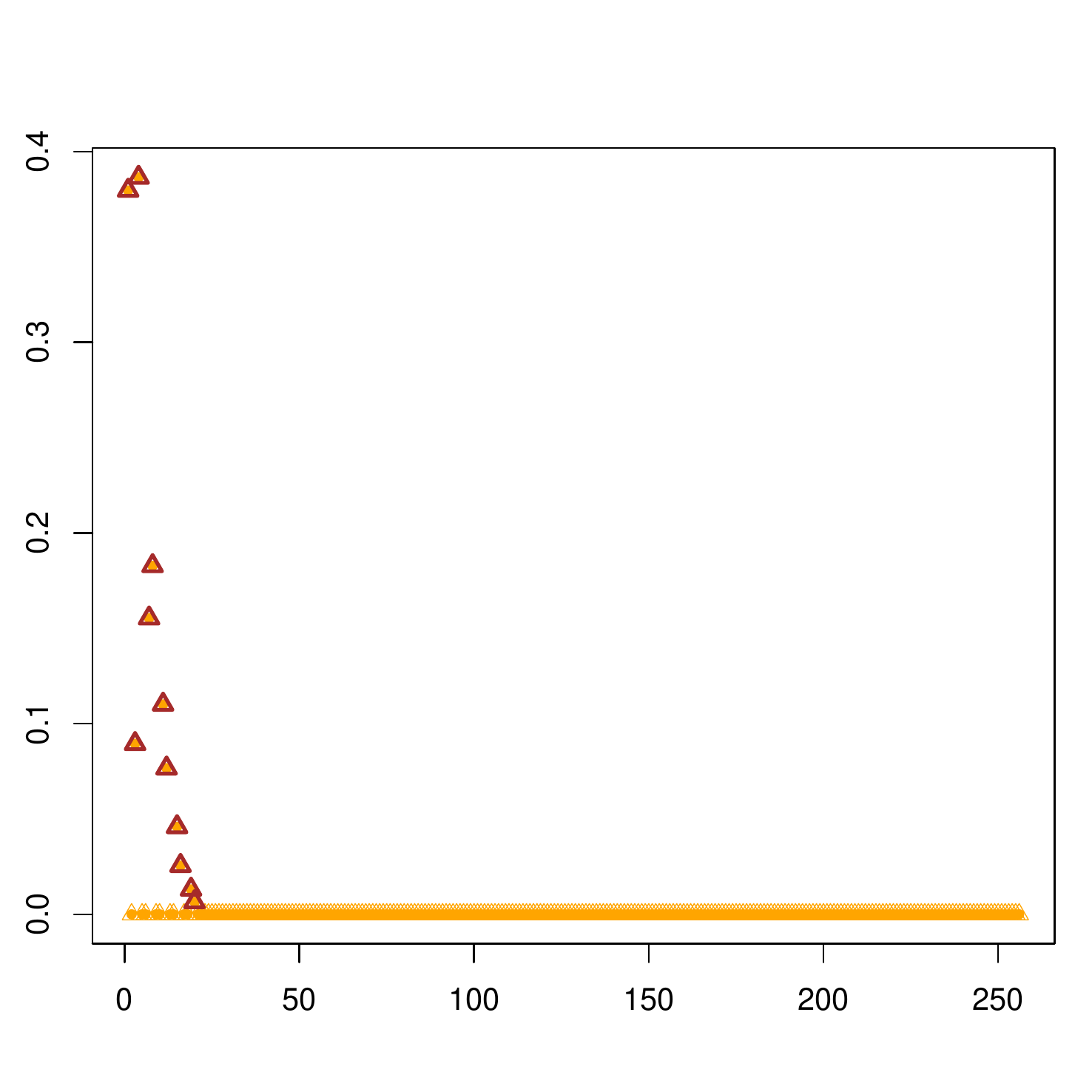}} \\
    \subfigure[Haar basis approximation]{\label{fig:stoch-c}\includegraphics[scale=0.3]{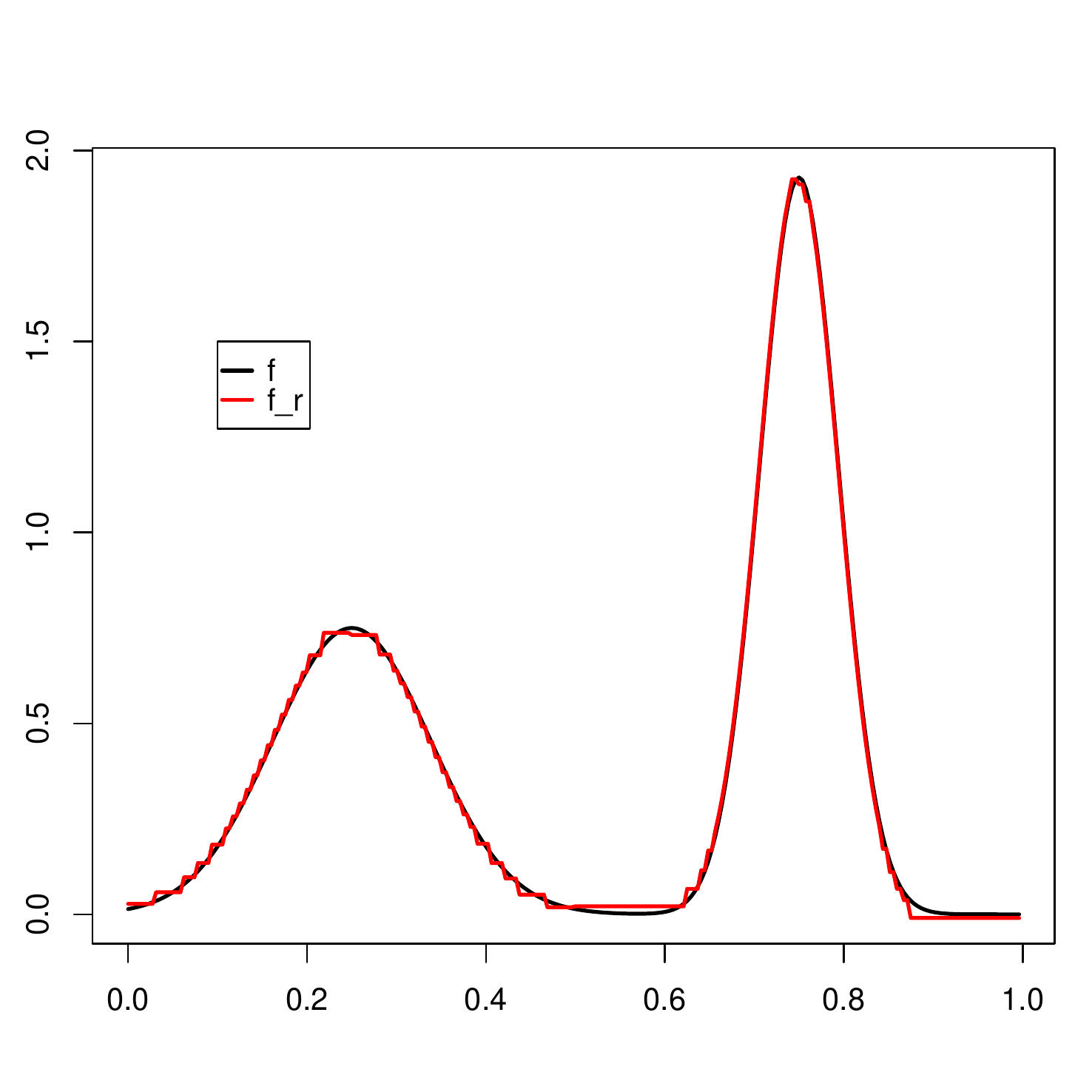}}
        \subfigure[Corresponding nonzero coefficients]{\label{fig:stoch-d}\includegraphics[scale=0.3]{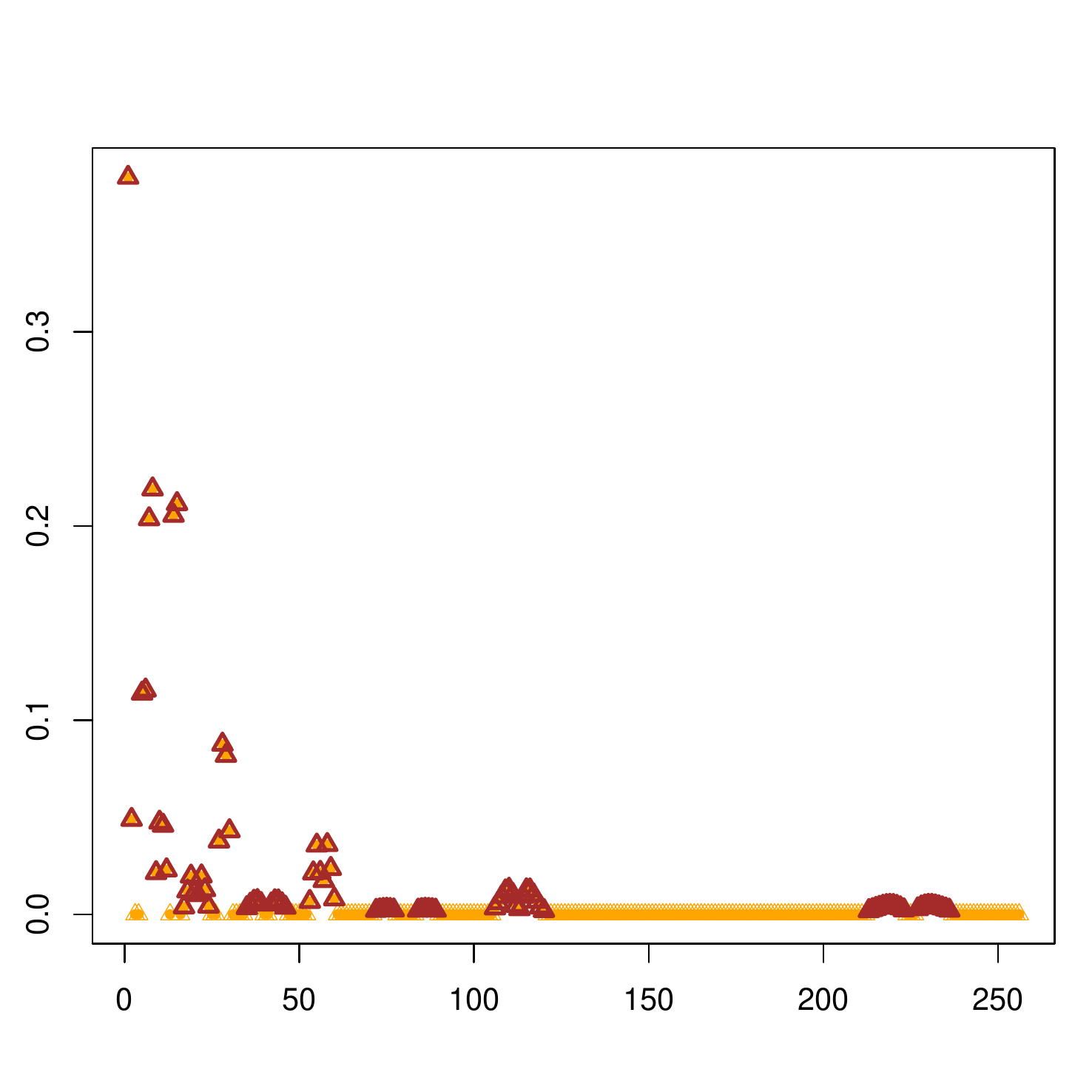}}
  \end{center}
  \caption{Sparsity of the mean function $f$ relative to the Haar and Fourier bases}
  \label{sparsity}
\end{figure}

To discuss the quality of the estimates given above relative to the mean $f$, we first investigate their properties relative to a truncated version of $f$ and obtain the desired results as a consequence. We motivate the truncation of $f$ below. Consider the  theoretical truncation level
\begin{equation}\label{theolevel} r_k =\sqrt{ \frac{ \sigma_k^2 + \sigma_\eps^2/m}{n} }
 z\left(\frac{\alpha}{2m}\right), \end{equation}
where $z(\alpha)$ is the quantile corresponding to a  $\mathcal{N}(0,1)$ random variable and $0 < \alpha < 1$ is a given number.  Notice that this is the population counterpart of the data based levels given in (\ref{trunc}) above. Define, for each $1\le k\le m$,
\[ \mu_{k}(r_k) = \mu_k \1\{ |\mu_k|\ge r_k\}
\] 
and we write
\[ \bar f_{(r)} (t)= \sum_{k=1}^m \mu_k(r_k)  \phi_k(t)\]
for a truncated version of $f$.  When the truncation $\bar f_{(r)}$ retains the main features of $f$,  it can be considered as the new target for inference. This is the approach we take in the sequel. As an illustrative example, we consider  the mean function 
\label{gaussian_mix}
\[ f(t) = 0.75 \exp\left\{ 64(t-0.25)^2 \right\} + 1.93 \exp\left\{-256(t-0.75)^2\right\}, \]
shown in black in Figures \ref{sparsity}(a)  and \ref{sparsity}(c). We project $(f(t_1), \ldots, f(t_m))' $ onto linear subspaces   in $\RR^m$ generated by the  Fourier and Haar basis functions, respectively, evaluated at $m=2^{8}=256$ equally distant points in $[0, 1]$.  Many of  the projection coefficients $ \mu_k = ({1}/{m})\sum_{j=1}^{m}f(t_{j})\phi_k(t_j)$ are   close to zero for both bases. We consider the truncation level $r_k$ given above with $\alpha = 0.05$, $n=400$, $\sigma_{\eps}^{2}=0.136$ and for $\sigma_{k}^{2}$ given by (\ref{var}) above 
 corresponding to the Brownian Bridge process with covariance function $\Gamma(s,t)=\text{min}(s,t)-st$. \\

\noindent Figures \ref{sparsity}(b)  and \ref{sparsity}(d) show $|\mu_{k}| \II_{\{|\mu_{k}| \geq r_{k}\}}$ versus their index, for the Fourier and Haar basis respectively. 
 The reconstruction $ \bar{f}_{r}(t)=\sum_{k=1}^{m} \mu_{k} \II_{\{|\mu_{k}| \geq r_{k}\}} \phi_{k}(t)$ is shown in red in Figures \ref{sparsity}(a)  and  \ref{sparsity}(c), and is very close to $f(t)$ in both cases. Notice that  only 11 coefficients are needed for this good reconstruction of $f$ via the Fourier basis, versus 92 via the Haar basis, as we reconstruct a differentiable function with differentiable and non-differentiable basis functions, respectively. Following standard terminology in non-parametric estimation, we refer to the fact that $f$ can be reconstructed well via a smaller subset of the given collection of the basis functions by saying that $f$ has a sparse representation relative to that basis. \\
 
 In what follows we show that the thresholded estimates introduced in the section above adapt to the sparsity of $f$. Of course, since $f$ is unknown, so is its sparsity relative to a
 given basis. Nevertheless, we show that our estimators adapt to this unknown sparsity in terms of their fit, and refer to these results as oracle inequalities. The type of oracle inequalities that we establish below  illustrate that the fit of our estimators depends only on the estimation errors induced by the  estimates of the  non-zero coefficients of a sparse representation of $f$ within a given basis. As our example indicates, and as is the case  in any non-parametric estimation problem, the overall quality of our estimator will further depend on the choice of the basis used for estimation. We will therefore complement the construction of our estimator with a basis selection step. We begin by stating our results for a given basis. \\

For both hard and soft threshold estimators we obtain estimation bounds on the fit at the observation points $t_j$. We formulate our results in terms of the empirical supremum norm  $ \| \ \|_{m,\infty}$  and $L_2$ norm $\| \ \|_{m,2} $ defined below. For any real function $g$, let
\begin{eqnarray*}
  \| g\|_{m,\infty} = \max_{1\le j \le m} | g(t_j)|; \ \  \ \ 
\| g\|_{m,2} = \sqrt{\frac1m\sum_{j=1}^m g^2(t_j) }.
\end{eqnarray*}

All theorems and results of this article require that $\EE[ |Z(t)|^2]<\infty$, for all $t$,  and that $\EE[\eps^2]<\infty$.  The next three theorems are proved for a given basis  and the desired probability $\alpha$. All estimates are based on the  threshold level $\wh r_k=\wh r_k(\alpha)$ given in (\ref{trunc}), for a user specified value of  $\alpha$.
The following result establishes oracle inequalities for the hard-threshold estimators. 
  Define
 \begin{eqnarray}\label{tech}
  \bar  r_k &=& \frac{ z(\frac{\alpha}{2m}) }{\sqrt{n} } \left\{ \sqrt{ \sigma_k^2 + \frac{\sigma_\eps^2}{m} } + 2\delta \right\},
 \end{eqnarray}
\noindent which differs by $2\delta$ from $r_k$ defined above in (\ref{theolevel}), for a quantity $\delta$ that is arbitrarily close to zero; this is needed for purely technical reasons, and for all practical purposes $\bar r_k$ and $r_k$ can be considered the same.  

\begin{thm}\label{Thm1} 
For all $m\ge 1$, $0<\alpha<1$ and $\delta>0$
\begin{eqnarray*}
\| \wh f_{(2 \wh r)} - \bar f_{(r)}\|_{m,\infty} &\le& 3\max_{1\le k\le m} \|\phi_k\|_\infty \sum_{k=1}^m
\bar r_k\1 \{ |\mu_k|\ge r_k\} \\
\| \wh f_{(2\wh r)} - \bar f_{(r)}\|_{m,2} &\le& 3\sqrt{ \sum_{k=1}^m \bar r_k ^2
\1\{ |\mu_k|\ge r_k\}},\\
\end{eqnarray*}
with probability at least $1 - \alpha$, as $n \rightarrow \infty$. 
\end{thm}

\medskip
The same conclusion holds for the soft-threshold estimator:\\

\begin{thm} \label{Thm2}
Let $\bar{r}_k$ be as in (\ref{tech}). For all $m\ge 1$, $0<\alpha<1$ and $\delta>0$
\begin{eqnarray*}
\| \widetilde f_{(2\wh r)} - \bar f_{(r)}\|_{m,\infty} &\le& 3\max_{1\le k\le m} \|\phi_k\|_\infty
\sum_{k=1}^m\bar
r_k \1\{ |\mu_k|\ge r_k\} \\
\| \widetilde f_{(2\wh r)} - \bar f_{(r)}\|_{m,2} &\le& 3\sqrt{ \sum_{k=1}^m
\bar r_k ^2 \1 \{ |\mu_k|\ge r_k\}},\\
\end{eqnarray*}
with probability at least $1 - \alpha$, as $n \rightarrow \infty$. 
\end{thm}

 Theorems \ref{Thm1} and \ref{Thm2} yield immediately results on the performance of these estimators relative to the untruncated $f$.  In particular, the following inequality holds with probability at least $1-\alpha$ as $n\to\infty$:
\begin{eqnarray}\label{fullf}
\|\wh g - f\|_{m,2} \le \| \bar f_{(r)} -f \|_{m,2} + 3\sqrt{ \sum_{k=1}^m \bar r_k^2 \1\{ |\mu_k|> r_k\} }
\end{eqnarray}
for both estimators $\wh g=  \wh f_{(2\wh r)}$ and $ \wh g=  \widetilde f_{(2\wh r)}$. This follows directly from the above results and the triangle inequality.
The first term $\| \bar f_{(r)}-f\|_{m,2}$ can be viewed as the approximation error or bias term, whereas the second term represents the estimation error or 
standard deviation term.  The bias term is unavoidable, and its size depends on the basis choice.  It suggests the need for an adaptive method, that would select the basis that is best suited for the unknown underlying mean function $f$. We discuss this in Section 2.4 below. \\

Theorems  \ref{Thm1} and \ref{Thm2} are novel type of oracle inequalities for thresholded estimators, as they guarantee the ``in probability", rather than ``on average",  performance of the estimator,  at any probability level of interest $0 < \alpha < 1$.  These properties hold for our estimates, as they are constructed relative to variable threshold levels that depend on $\alpha$. To the best of our  knowledge, such results are new in the functional data context. They are also new  in the general non-parametric settings, where a  more traditional way to state the oracle properties of the estimators is in terms of the expected mean squared error, see, for instance, Donoho and Johnstone (1995, 1998), Wasserman (2006),   Tsybakov (2009) and the references therein.  For completeness, we also give an assessment  of our estimates in terms of the expected mean squared error in Theorem \ref{Thm3} below,
 which restates Theorem \ref{Thm1} in terms of expected values. To avoid technical clutter, we consider the toy estimator $\wh f_{(2r)}$ in lieu of $\wh f_{(2\wh r)}$. Recall the notation 
\[\label{theolevel1} r_k = \sqrt{ \frac{\sigma_k^2}{n} + \frac{\sigma_\eps^2}{mn} } \cdot 
 z\left(\frac{\alpha}{2m}\right).\]

\begin{thm} \label{Thm3}
For all $m\ge1 $, $0<\alpha<1$ and $\delta>0$, we have
\begin{eqnarray}\label{gen}
 \EE \| \wh f_{(2r)}- \bar f_{(r)} \|_{m,2}^2
&\le& 2\sum_{k=1}^m \left(\frac{\sigma_k^2}{n} + \frac{\sigma_\eps^2}{mn}+ 4r_k^2\right)  I\{ |\mu_k|>r_k\} \\
&&+ \ 4  \sum_{k=1}^m \EE\left[  \left( r_k^2 +    (\wh \mu_k-\mu_k) ^2  \right) I\{|\wh \mu_k-\mu_k|> r_k\}\right]. \nonumber 
\end{eqnarray}
In particular, if $Z$ is a Gaussian process and $\eps_{ij}$ are Gaussian random variables,
\begin{eqnarray}\label{normal}
 \EE \| \wh f_{(2r)}- \bar f_{(r)} \|_{m,2}^2
&\le& 2\sum_{k=1}^m \left(\frac{\sigma_k^2}{n}+\frac{\sigma_\eps^2}{mn} 
 + \  4r_k^2 \right)  I\{ |\mu_k|>r_k\} \\
&& +  \ \frac{4\alpha}{m} \sum_{k=1}^m\left ( r_k^2+\frac{\sigma_k^2}{n}+\frac{\sigma_\eps^2}{n m} \right). \nonumber 
\end{eqnarray}
\end{thm}
\medskip

Theorem \ref{Thm3} shows that the expected mean squared error of our estimator also adapts to the unknown sparsity of $f$, as indicated by the first term in either inequality 
(\ref{gen}) or (\ref{normal}). The second term in these inequalities is essentially an average of the quantities that constitute the first term. This  is more evident from  the closed form expression (\ref{normal}), and shows that this second term is negligible relative to the first one, especially for small values of $\alpha$.

\subsection{Data adaptive basis selection}

The results of the previous section make it  clear that the basis choice influences 
both the bias and the variance of our estimates; the type of basis one uses for the fit can be regarded as the tuning parameter of our estimation procedure.  We give below a data adaptive procedure of selection and show in Theorem \ref{ds} below that the estimator based on the selected basis behaves essentially as if the best basis for approximating the unknown $f$ was known in advance. \\

We select the  basis  via a cross-validation (data-splitting) technique, by randomly dividing the $n$ discretized curves $\{ (Y_{ij},t_j),\ 1\le j\le m\}$ in two equally sized groups.  The first sample 
$\{ (Y_{ij},t_j),\ 1\le j\le m,\ i\in I_1\}$
is used for constructing various estimates, say $\wh g_\ell$, $1\le \ell\le L$, based on various bases, choices of $\alpha$ and thresholding methods (hard and soft). The second sample (hold-out or validation sample) $\{ (Y_{ij},t_j),\ 1\le j\le m,\ i\in  I_2\}$ is used to select the optimal estimate $\wh g= \wh g_{\wh\ell}$ that minimizes the empirical risk
\[ \frac{1}{n_2} \sum_{i\in I_2} \frac 1m \sum_{j=1}^m \{ Y_{ij}- \wh g_\ell (t_j) \}^2 \]
over $\ell=1,\ldots, L$. Here $I_2$ is the index set for the curves that are set aside to evaluate the estimators $\wh g_\ell$ and $n_2=|I_2|$ is its cardinality.

\begin{thm}\label{ds} Assume that $\tau_{\eps,p}\equiv\EE |\eps|^p<\infty$ and
\[\bar \tau_{Z,p} \equiv \frac1m \sum_{j=1}^m\EE | Z_i(T_j)|^p <\infty \] for some $p>2$.
The minimizer $\wh g= \wh g_{\wh \ell}$ satisfies
\begin{equation}\label{sel}
\EE\|\wh g-f\|_{m,2}^2 \le
2 \left[ \min _ {1\le \ell\le L}  \EE\|\wh g_\ell -f\|_{m,2}^2 + \frac1n +  
L \frac{  2 ^{p/2}C_p}{n} \left(\tau_{\eps,p}+\sigma_\eps^p+ \bar\tau_{Z,p}+\bar\tau_{Z,2}^{p/2} \right) \right],
\end{equation} for some constant $C_p\le 7.35p/\max\{1,\log(p)\}$.
\end{thm}

\medskip
{\it Remark.} Theorem \ref{ds} requires that the process $Z(t)$ and the random error $\varepsilon$ have moments strictly larger than 2, which is still a very mild assumption.\\

The last term in the right hand side of (\ref{sel}) is of order $1/n$, making the sum of the last two terms of order $1/n$. This can be regarded as the price to pay for using a data adaptive procedure to select the appropriate basis. The factor 2 multiplying the right hand side of (\ref{sel}) can be reduced to $1 + \beta$ at the cost of increasing the last two terms on the right by a factor proportional to $1/\beta$, for $\beta >0$ arbitrarily close to zero. To avoid notational clutter we opted for using the constant 2. Therefore, Theorem \ref{ds}  shows that the basis selection process yields an estimate that is essentially as good as the best estimate on the list, in terms of expected squared error. Since which is best cannot be known in advance, as $f$ is unknown, the result of Theorem \ref{ds} can also be regarded as an oracle inequality. 

\subsection{Confidence bands}

In this section we will construct confidence bands for $f$ that are uniform over the parameter space. We begin with the  confidence band based on a hard  threshold
estimator given below. Set \[ \widetilde r_k =\frac{z(\alpha/2m)}{\sqrt{n} } ( S_k+ 3\delta),\]
and notice that it differs by $\delta$ from $\wh r_k$ given in (\ref{trunc}) above. This is again needed for technical reasons, as in practice $\delta$ can be set to zero.
\begin{thm}\label{Thm5}
(1)  For all $m\ge 1$, $0<\alpha<1$ and $\delta>0$ 
\begin{eqnarray*}
\left\{ \wh f_{(\wh r)}(t_j) \pm \sum_{k=1}^m (3 \widetilde r_k) |\phi_k(t_j )| \1\{ |\wh\mu_k|>
\wh r_k\},\ 1\le j\le m\right\}
\end{eqnarray*}
contains $\{ \bar f_{(2 \bar r)}(t_j),\ 1\le j\le m\}$ with probability at least $1-\alpha$, as $n\to\infty$. \\

(2)  Moreover, if all non-zero coefficients $\mu_k$ exceed $2 \bar r_k$, the band can be made smaller by a factor 3:
\begin{eqnarray*}
\left\{ \wh f_{(\wh r)} (t_j) \pm \sum_{k=1}^m \widetilde r_k |\phi_k(t_j )| \1\{ |\wh\mu_k|>\wh 
r_k\},\ 1\le j\le m\right\}
\end{eqnarray*}
contains $\{ \bar f_{(2\bar r)}(t_j),\ 1\le j\le m\}$ with probability at least $1-\alpha$, as $n\to\infty$.
\end{thm}

We obtain similar results for the soft-threshold estimator.\\

\begin{thm}\label{Thm6}
For all $m\ge 1$, $0<\alpha<1$ and $\delta>0$
\begin{eqnarray*}
\left\{ \widetilde f_{(\wh r)} (t_j) \pm \sum_{k=1}^m (2\widetilde r_k)| \phi_k(t_j)| \1\{
|\widehat \mu_k|>\wh r_k\},\ 1\le j\le m\right\}
\end{eqnarray*}
contains $\{ \bar f_{(2\bar r)}(t_j),\ 1\le j\le m\}$ with probability at least $1-\alpha$, as $n\to\infty$.\\
\end{thm}

Lemma \ref{lem} in the Appendix below and the remark following it  show that the bands have asymptotic probability $1-\alpha$ {\it  uniformly}  in the parameter $f$ or, equivalently, in the parameter $(\mu_1,\ldots,\mu_m)$.  This rules out the possibility of  exhibiting, for each $n$, a ``bad" parameter value $(\mu_1,\ldots,\mu_m)$ for which the coverage probability is much smaller than $1-\alpha$.  This would have been the case had we based our construction on the limiting distribution  of the truncated estimators $\wh f$ or $\widetilde f$, when the resulting confidence bands cannot be expected to be uniform, as pointed  out by, for instance,  Genovose and Wasserman (2008) and Wasserman (2006).\\

Typically, the price to pay for having uniform confidence bands is the width of the band, which is necessarily larger than the width of a pointwise  band, as a uniform band  needs to cover all unfavorable cases. However, if we restrict the space of the parameters over which we require uniformity to spaces containing only   those $\mu_k$ that are  above a small threshold, the width of our bands can be made significantly narrower, as in {\it (2)} of Theorem \ref{Thm5} above.  Theorems \ref{Thm5} - \ref{Thm6} above show that 
the average width of the bands is a multiple (1, 2 or 3), depending on the estimator and type of uniformity, of 
\[ \frac1m \sum_{j=1}^m \left( \sum_{k=1}^m \widetilde r_k |\phi_k(t_j )| \1\{ |\wh\mu_k|\ge \wh r_k\} \right) = 
\sum_{k=1}^m \widetilde r_k  \left(\frac1m \sum_{j=1}^m|\phi_k(t_j )|\right) \1\{ |\wh\mu_k|\ge \wh r_k\} .
\]

\noindent {\it Remark.} From the expression above it is clear that the bands and their width adapt to the unknown sparsity of $f$, as only  the coefficients $\wh \mu_k$
above the given threshold  $\wh r_k$ contribute to the band and they, in turn, estimate the true coefficients above a certain  threshold, which reflect the sparsity of $f$. Since sparsity is relative to a given basis, in practice the construction of a confidence band is based on the best basis selected from a library of bases, as discussed in Section 2.4 above. Finally, in order  to obtain {\it uniform}  bands of reasonable width,  we constructed confidence bands for the surrogate $\bar f_{2r}$ of $f$, as  advocated by  Genovese and Wasserman (2008) for bands in standard nonparametric regression models based on non-functional data; this surrogate is based on the best selected basis  and will capture the main features of $f$, as illustrated in Figure 1 of Section 2.2.

\bigskip

\section{Numerical results}

\subsection{Simulation design}

We conducted our simulations for a combination of types of stochastic processes, stationary and non-stationary, and differentiable and non-differentiable mean functions. Specifically, we consider two stationary processes, AR(1) and ARIMA(1,1), and two non-stationary processes, the Brownian Bridge (BB) and the Brownian Motion (BM) on [0,1]. We consider the two mean functions: $f(t) = c_1 \exp{ \lbrace-64(t-0.25)^2\rbrace} + c_2 \exp{\lbrace-256(t-0.75)^2\rbrace}$, referred to in the sequel as Signal 1, and $f(t) = c_3\II_{0.35<t<0.375}+c_3\II_{0.75<t<0.875}$, referred to as Signal 2.  The constants $c_1 - c_4$ will be varied to achieve various desired signal to noise ratios.\\

For our simulations we considered  two popular families of bases, Fourier and Haar, each known to have good approximation properties for functions in $L^2([0,1])$ belonging to general smoothness classes, e.g. Sobolev classes.  Both bases share the orthonormality property $ \sum_{j=1}^m \phi_k(t_j) \phi_{k'}(t_j)= m\1{\{k=k'\}}$, 
for $1\le k,k'\le m$, see, e.g. Tsybakov (2009) for an argument and a more detailed discussion of smoothness classes.  Any other bases with this property can be considered, and the qualitative and quantitative points we illustrate here will remain essentially the same.

\subsubsection{Simulation scenarios}

\noindent We simulated $n$ curves for each of the eight combinations (signal, stochastic process)
above. Each curve $ 1 \leq i \leq n$  is observed at $m $ equally spaced points $t_j \in [0, 1]$  and the observations follow model (\ref{model}), 
\[ Y_{ij}=f(t_{j})+Z_{i}(t_{j})+\eps_{ij},\]
\noindent for $ 1 \le i \le n$ and $ 1 \le j \le m$. The measurement errors  $\eps_{ij} \sim N(0,\sigma_{\eps}^{2})$ are i.i.d. across $i$ and $j$. The parameters for simulating the AR(1) and ARIMA(1,1) processes are chosen in order to achieve the following equivalences: \\
$\underbrace{\text{median}}_{t}$  \{ Var [ Brownian Bridge (t)] \} = Var \{AR(1)\} \\
$\underbrace{\text{median}}_{t}$ \{ Var [ Brownian Motion (t)] \} = Var \{ARIMA(1,1)\} \\

\noindent This facilitates  comparison between processes of different natures.   Next, the variance of the measurement error $\sigma_{\varepsilon}^{2}$ is chosen so that we have two cases: $\sigma^{*}=1$ and $\sigma^{*}=10$, where
\begin{equation}\label{star} \sigma^{*} =  \dfrac{\text{var} [Z(t)]}{\sigma^2_{\varepsilon}}.  \end{equation}
When $\sigma^* = 1$ the variability of the measurement error is the same as that of the stochastic process, whereas for  $\sigma^* = 10$ the measurement errors become essentially negligible. Figure \ref{fig:stoch_M} below shows, respectively,   realizations from  each of the stochastic process with mean corresponding to Signal 1 and  added noise corresponding to $\sigma^* = 1$  and 10, respectively. \\

\begin{figure}[]
  \begin{center}
    \subfigure[AR(1), $\sigma^{*}=10$]{\label{fig:stoch-a}\includegraphics[scale=0.21]{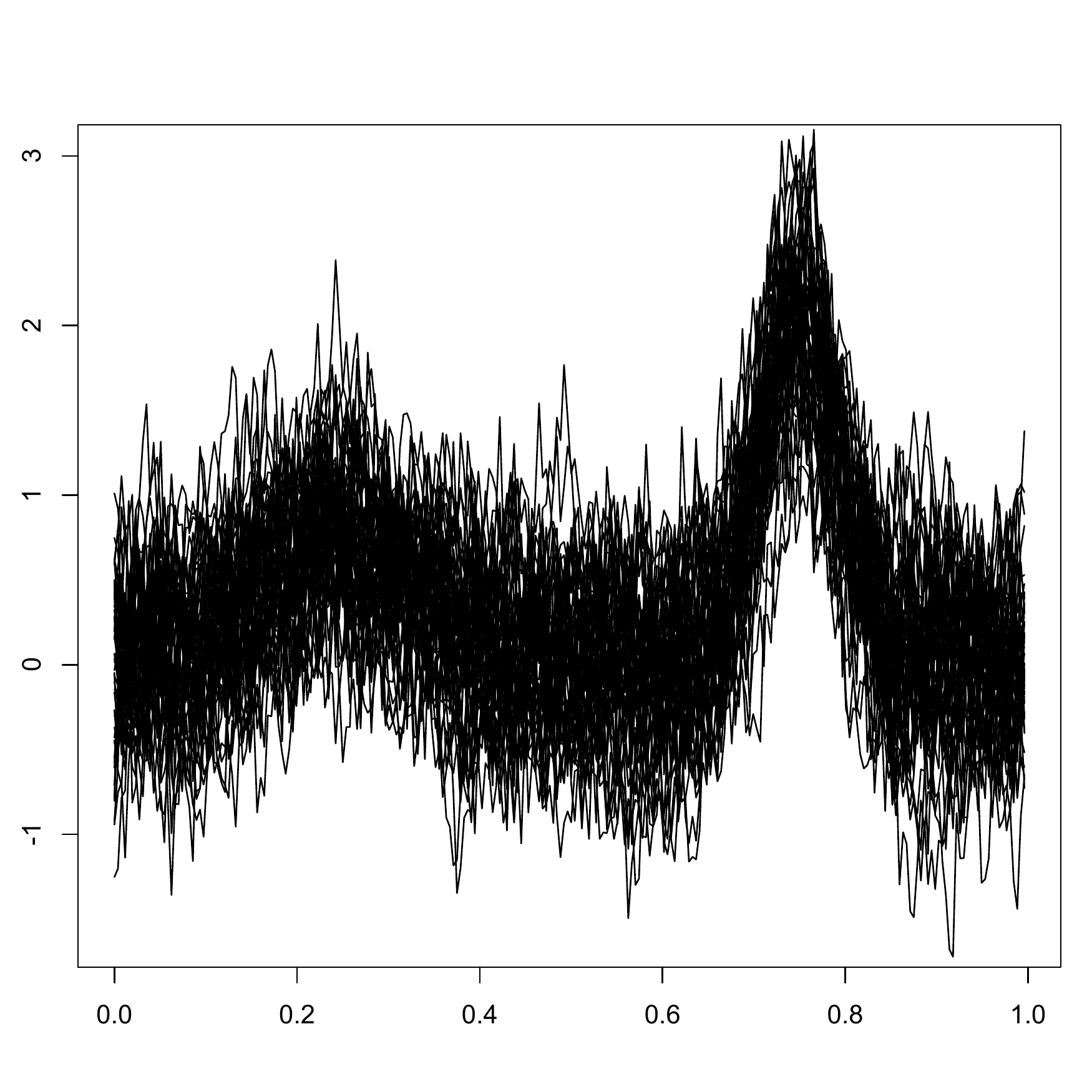}}
    \subfigure[AR(1), $\sigma^{*}=1$]{\label{fig:stoch-b}\includegraphics[scale=0.21]{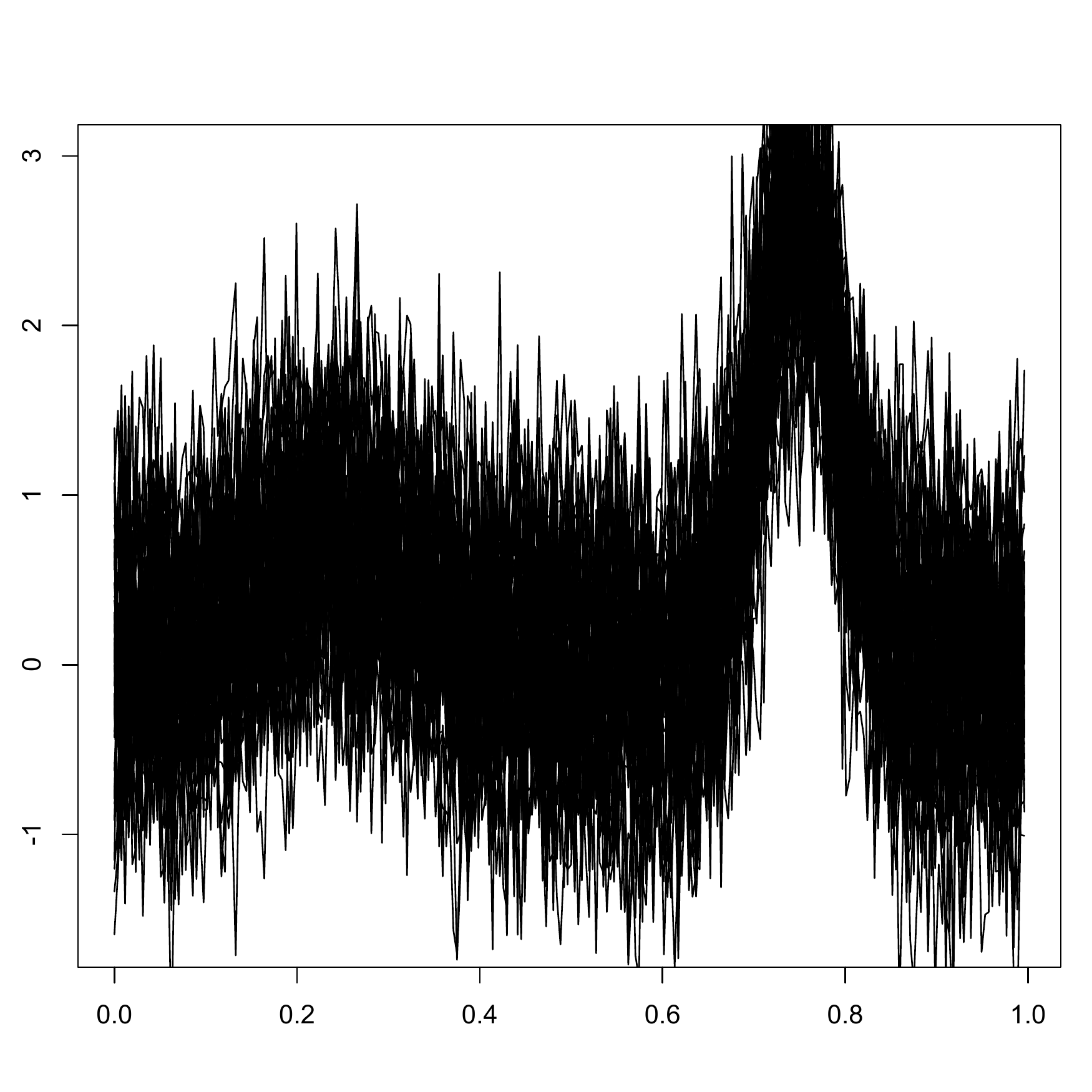}} 
    \subfigure[BB, $\sigma^{*}=10$]{\label{fig:stoch-c}\includegraphics[scale=0.21]{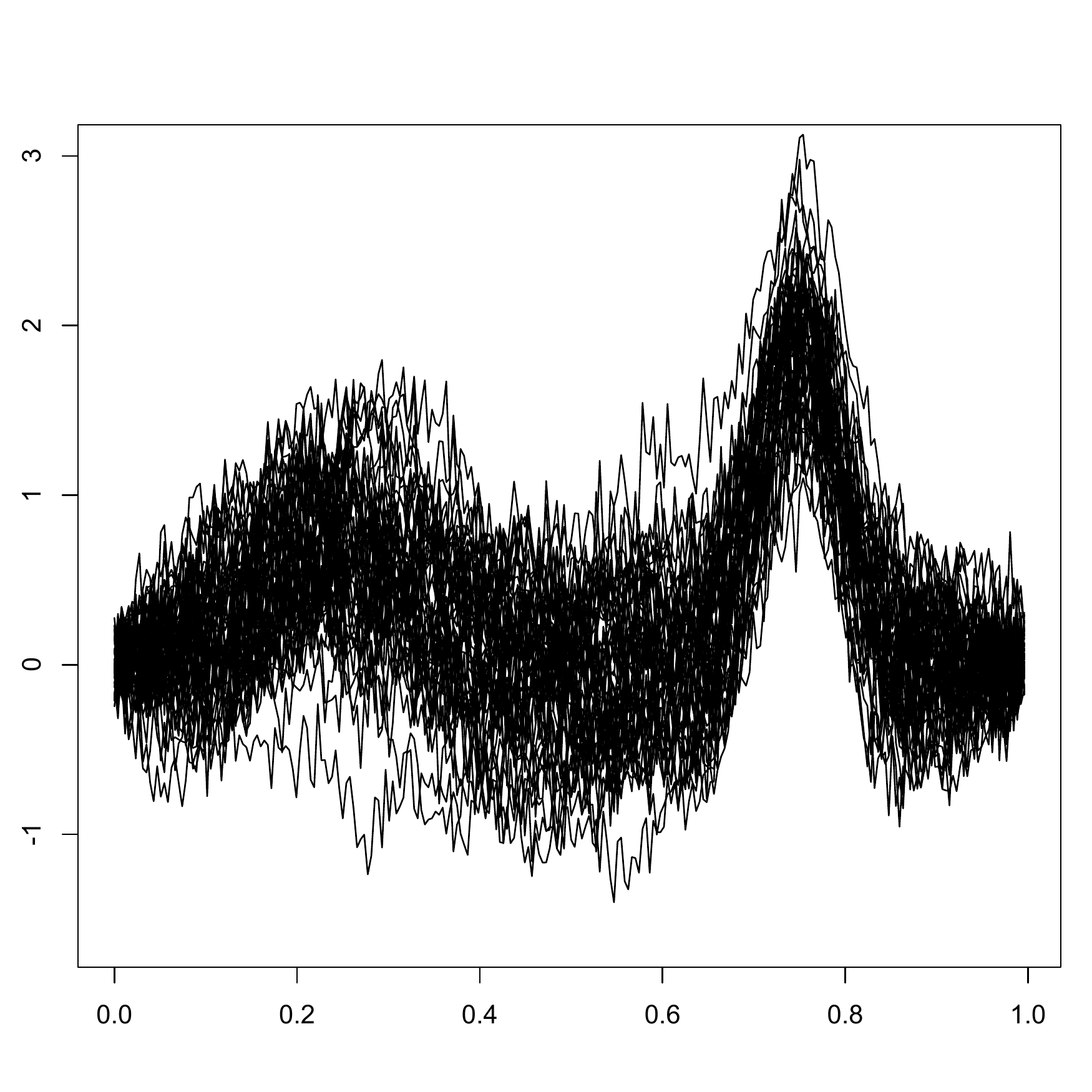}}
        \subfigure[BB, $\sigma^{*}=1$]{\label{fig:stoch-d}\includegraphics[scale=0.21]{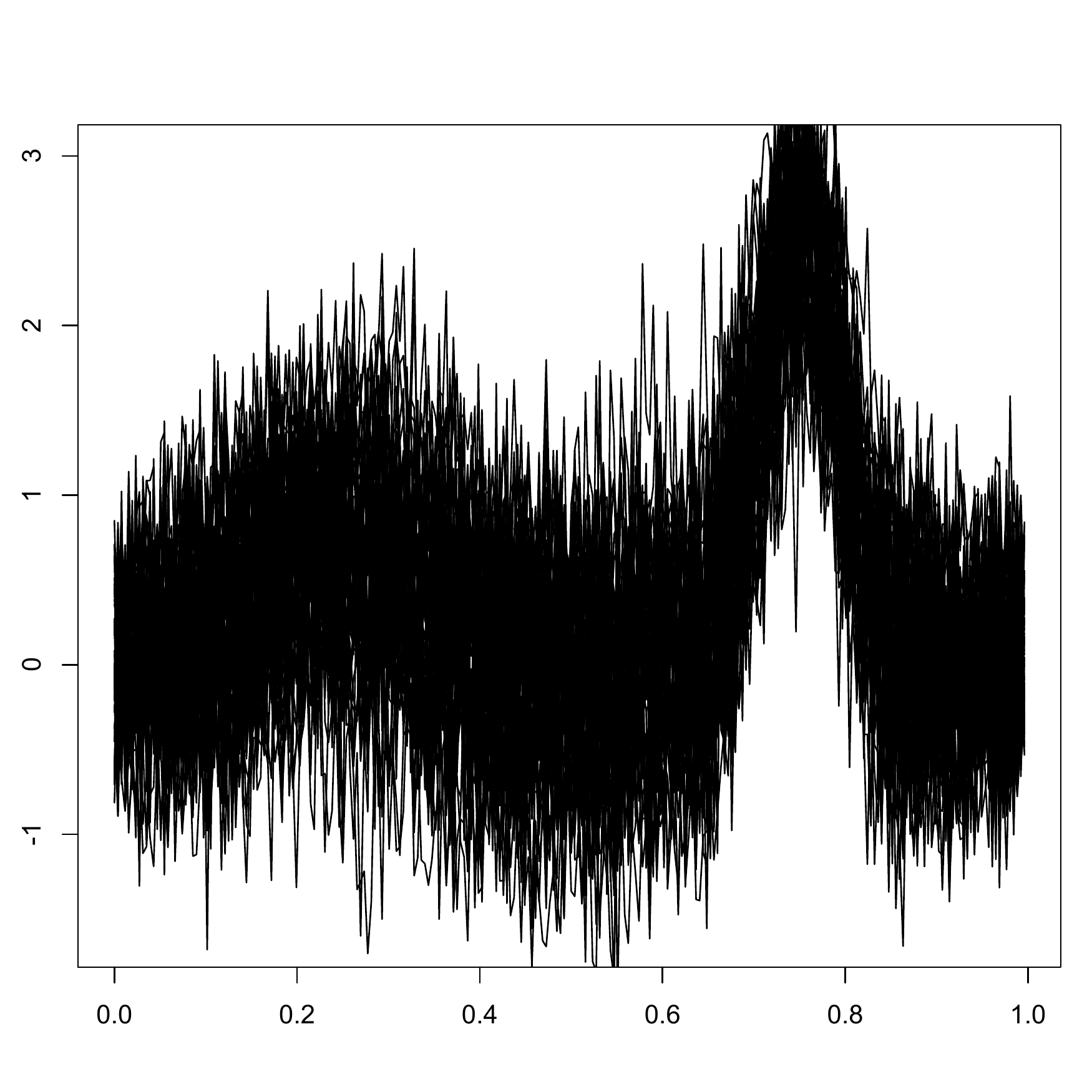}} \\
      \subfigure[ARIMA(1,1), $\sigma^{*}=10$]{\label{fig:stoch-a}\includegraphics[scale=0.22]{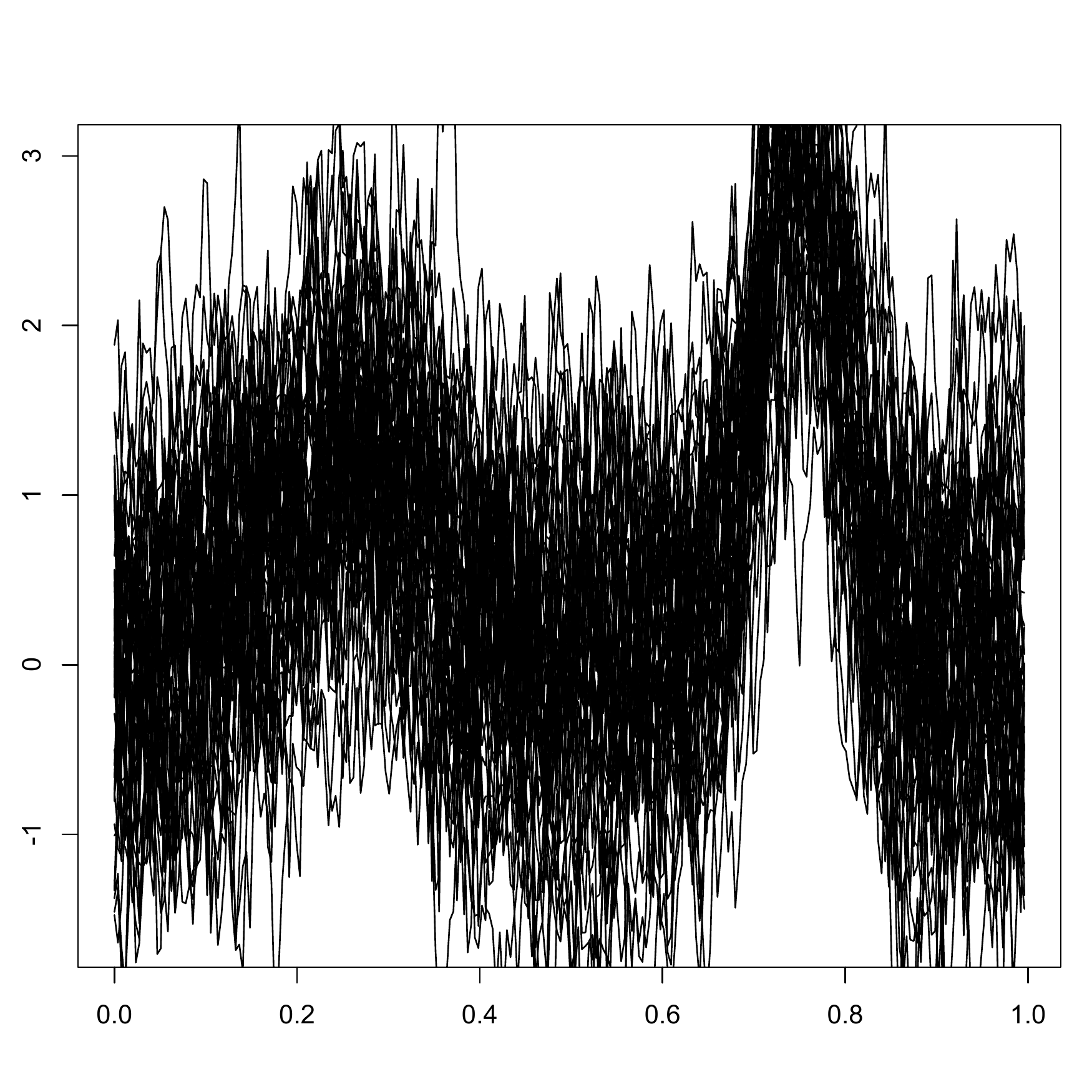}}
    \subfigure[ARIMA(1,1), $\sigma^{*}=1$]{\label{fig:stoch-b}\includegraphics[scale=0.21]{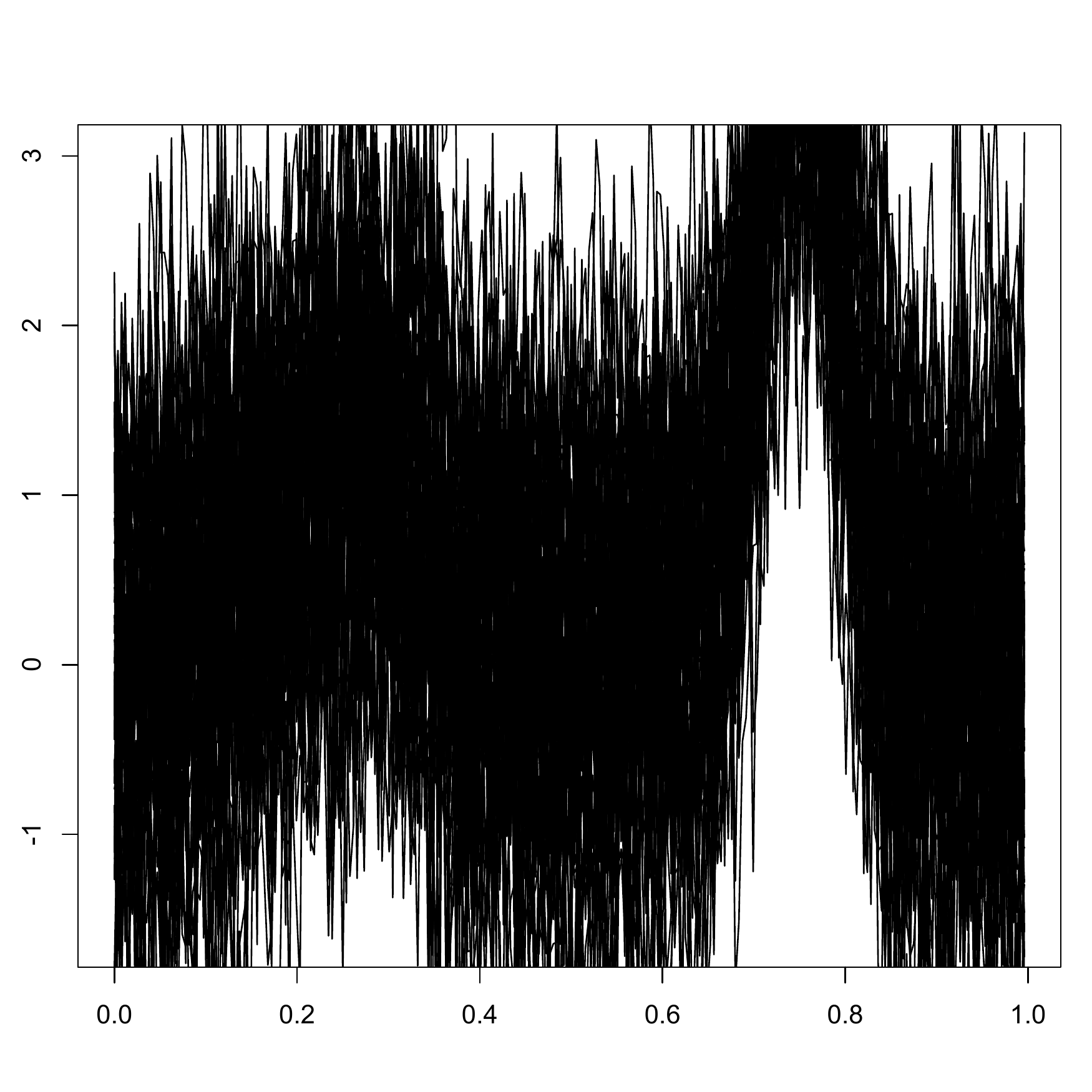}} 
    \subfigure[BM, $\sigma^{*}=10$]{\label{fig:stoch-c}\includegraphics[scale=0.21]{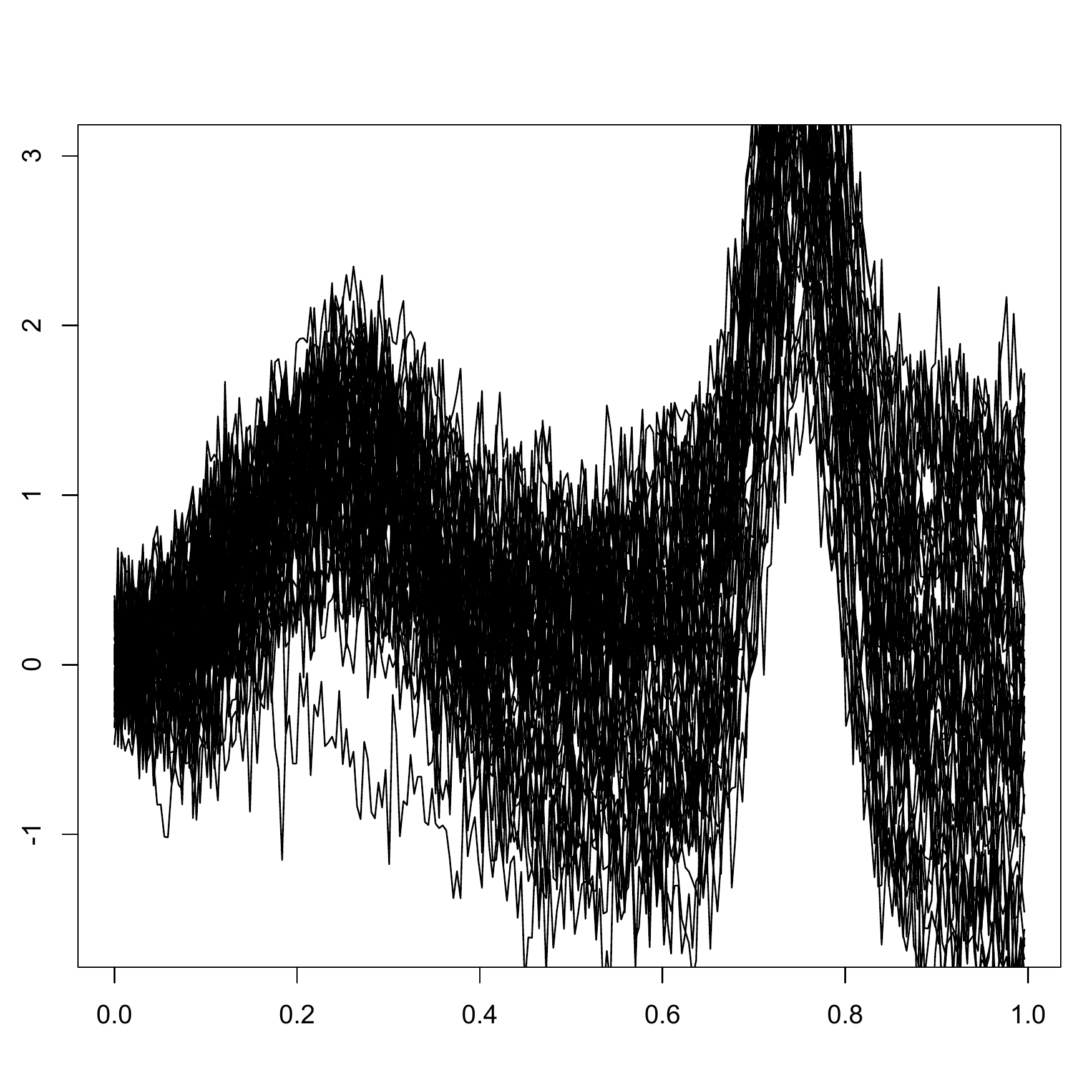}}
        \subfigure[BM, $\sigma^{*}=1$]{\label{fig:stoch-d}\includegraphics[scale=0.21]{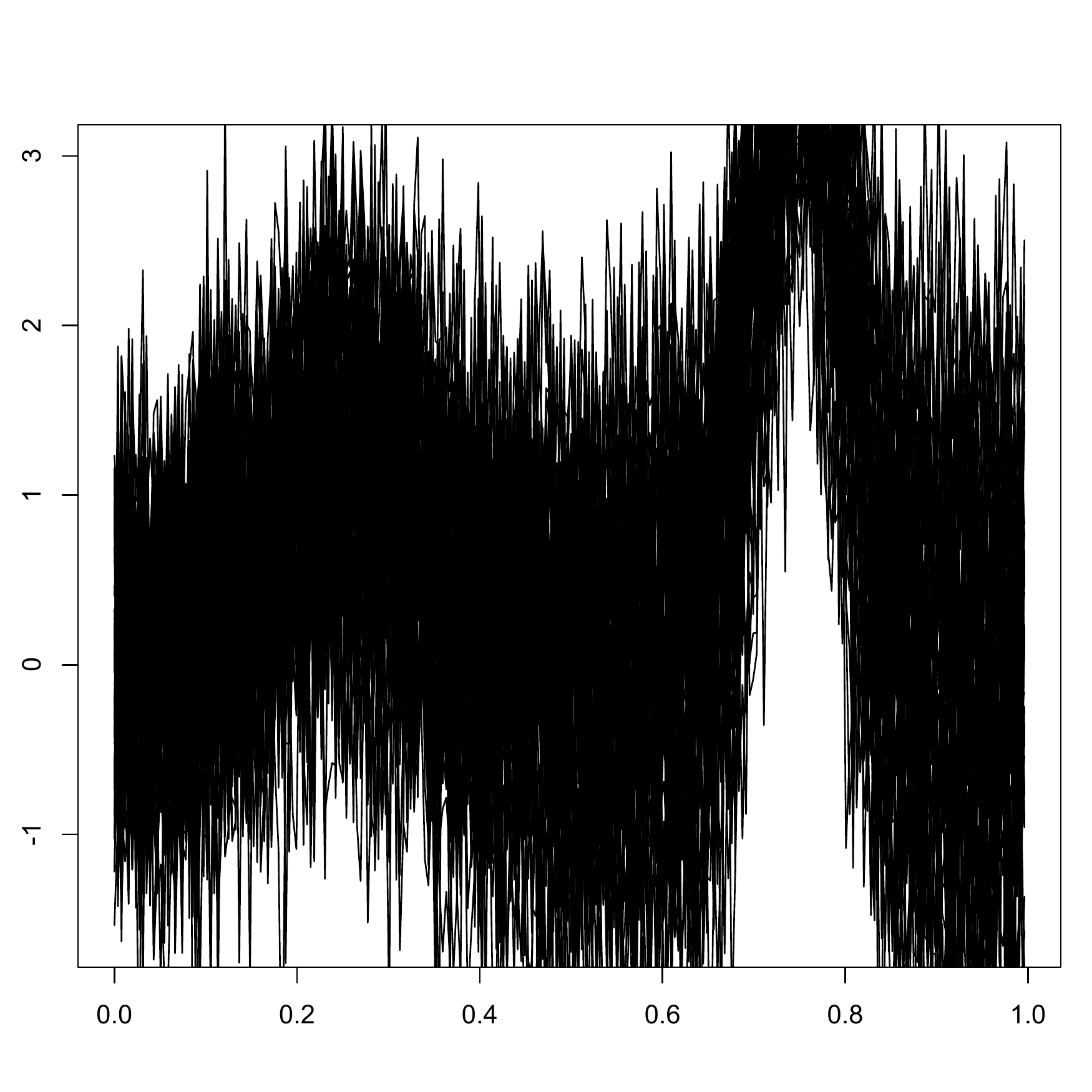}}
  \end{center}
  \caption{ Plots of Signal 1 + AR(1)/BB + Noise (top row) and  of Signal 1 + ARIMA(1,1)/BM + Noise (bottom row), $n =50$, $m = 256$, SNR = 4.25}
  \label{fig:stoch_M}
\end{figure}

\noindent We conducted simulations for different values of the signal-to-noise ratio (SNR). Since 
the process $Z(t)$ is assumed independent of the measurement error, we define as a measure of the noise $\left({\text{Var} [Z(t)]  + \sigma_{\eps}^{2} }\right)^{1/2} $ and the  signal-to-noise ratio to be 
$ \text{SNR}  = \text{Range}[f] / \left( \text{Var} [Z(t)]  + \sigma_{\eps}^{2} \right)^{1/2}$,
where $\text{Range}[f]=|\max_t  f(t) - \min_t   f(t)|$,  $t \in [0,1]$.

\subsection{Simulation results: the fit of the estimates}

In this subsection we considered $n = 400$ and $m = 256$; we took significantly lower values of $m$ and $n$ in the next subsection.  We contrast the quality of the fit of our estimates with the simplest estimate, the ensemble average of the observations $Y_{ij}$ and with estimates obtained via 7 other methods previously proposed and studied in the literature. The first three are obtained by applying, respectively, the following smoothing methods to the entire data set, containing all $n$ curves: (1) Linear polynomial kernel smoothing (Local Poly) with a global bandwidth, suggested by, among others, M\"uller (2005), Yao, M\"uller and Wang (2005), M\"uller, Sen and Stadtm\"uller (2006), Yao (2007).  We use a plug-in bandwidth adapting the method developed  by  Ruppert, Sheather and Wand (1995) to our case.
We obtain the estimated bandwidth $\wh{b}_{plug}$ using \texttt{dpill} in \texttt{R}, and the estimate $\wh{f}_{\text{locpoly}}(t)$ using \texttt{locpoly} in \texttt{R}; (2) 
Nadaraya-Watson kernel  smoothing (NWK) with a global bandwidth,  discussed in a functional data setting by, e.g., Yao (2007). 
We used a  Gaussian kernel, and a grid of possible bandwidth choices $G=[0.002, 0.004, 0.008, 0.01, 0.02, 0.04, 0.08, 0.1]$. We computed the estimate $\wh{f}_{NWK}(t)$ using each bandwidth in the grid $ b \in $ G. We report the results for the bandwidth $b \in$ G for which we obtained the smallest EMSE($\wh{f}_{NWK}$);
(3)  Smoothing splines, as suggested  by, for instance,  Rice and Silverman (1991). We used order 4 B-splines basis functions $B_{k}(t)$ with a knot placed at each design time point $t_{j}$, and the square of the second derivative of $f(t)$ as the roughness penalty. The tuning parameter in the penalty term  is chosen by generalized cross-validation,  leaving one curve out at a time. We implemented the method using  {\texttt{smooth.spline}} in {\texttt{R}}. \\

For the last four of the methods used for comparison we estimate $f(t)$ be averaging smoothed versions of the individual trajectories.  The reconstructions of the individual curves were  performed using:  (5) Linear polynomial kernel (Global kernel) smoother with a global bandwidth;  (6) Linear polynomial kernel smoother (Local kernel) with a local bandwidth, where the bandwidths are found using the plug-in algorithm proposed in Seifert, Brockman, Engel and Gasser (1994);   (7) B-splines regression with roughness penalty;  (8) Fourier expansion regression with roughness penalty, as discussed in Ramsay and Silverman (2005), Chapter 5.\\

We contrast the estimates above with our estimates.  We consider hard threshold estimates (HT) obtained by truncating the least squares estimates either at levels $\wh r_k$, for each $k$, and denote the resulting estimate
by HT($r$), or at levels $2\wh r_k$, to obtain HT($2r$), for $\wh r_k$ given in (\ref{trunc}) above, for $\delta = 0$.  We have also conducted extensive simulation experiments for the soft-thresholded estimates ST($r$) and ST($2r$), and in all cases we obtained inferior results to those obtained for the hard-thresholded estimators and for space limitations we do not report them in what follows. These results were expected, as the soft-thresolded estimator shrinks the least squares coefficients, and would need to be followed by a re-fitting step, which essentially amounts to the hard thresholding procedure we analyze below.  \\

\begin{table}[h]
\tiny
\begin{center}
\caption{EMSE results for Signal 1 for BB and AR(1)}
\begin{tabular}{l c c c c}
  \hline
 $\sqrt{EMSE}\times 10^{-6}$  &    
  \multicolumn{2}{c}{Brownian Bridge}
    &
   \multicolumn{2}{c}{AR(1)}\\

 [$\sqrt{MEDMSE}\times 10^{-6}$]& $\sigma^{\star} = 1$ & $\sigma^{\star} = 10$& $\sigma^{\star} = 1$ & $\sigma^{\star} = 10$\\
   \hline
\textit{\underline{Fourier Basis}}\\
OLS & 29582 [27580]& 21104 [18217] & 30767 [30398]& 22767 [22415] \\
HT(r) & 18429 [15723]& 17544 [14446] & 16928 [16072]& 15989 [15195] \\
HT(2r) & 20231 [17721]& 18334 [15441] & 23894 [23767]& 21448 [19651] \\

\textit{\underline{Haar Basis}}\\
OLS & 29493 [27537]& 21092 [18208] & 30672 [30302]& 22749 [22389] \\
HT(r) & 34820 [33551]& 23808 [21495] & 38204 [38064]& 27852 [27449] \\
HT(2r) & 48011 [47158]& 48879 [53945] & 61629 [61684]& 52427 [52232] \\

\textit{\underline{Pooled Curves}}\\
Local Poly &22769 [20274]& 20557 [17584] & 22673 [22075]& 20479 [20073] \\
NWK  & 23186 [20790]& 20271 [17289] & 22848 [22279]& 20594 [20153] \\
Smoothing Splines & 21455 [18801]& 20186 [17308] & 20794 [20263]& 19408 [19069] \\
Ensemble Ave & 29493 [27537]& 21092 [18208] & 30672 [30302]& 22749 [22389] \\

\textit{\underline{Curve - by - Curve}}\\
Global Kernel & 47709 [46877]& 23165 [20647] & 31641 [31123]& 20451 [20095] \\
Local Kernel & 36161 [34809]& 21677 [19036] & 27824 [27261]& 20548 [20152]\\
B-splines Regression & 38532 [37384]& 20181 [17291] & 21413 [20899]& 20897 [20531] \\
Fourier Regression & 39543 [38316]& 31905 [30200] & 38309 [38091]& 30359 [30167] \\
\hline
\end{tabular}
\label{Tab:BB_SinCos}
\end{center}
\end{table}

\begin{table}[h]
\tiny
\begin{center}
\caption{EMSE results for Signal 1 for BM and ARIMA(1,1)}
\label{B13}
\begin{tabular}{l c c c c}
  \hline
  $\sqrt{EMSE}\times 10^{-6}$ &    
  \multicolumn{2}{c}{Brownian Motion}
    &
   \multicolumn{2}{c}{ARIMA(1,1)}\\

[ $\sqrt{MEDMSE}\times 10^{-6}$]& $\sigma^{\star} = 1$ & $\sigma^{\star} = 10$& $\sigma^{\star} = 1$ & $\sigma^{\star} = 10$\\
   \hline
\textit{\underline{Fourier Basis}}\\
OLS & 50959 [44662]& 37958 [29482] & 53549 [52989] & 41562 [40893]\\
HT(r) & 35388 [26045]& 34133 [24780] & 31356 [30346] & 30491 [29488]\\
HT(2r) & 37794 [29300]& 35322 [24835] & 42625 [41283] & 44088 [39450]\\

\textit{\underline{Haar Basis}}\\
OLS & 50815 [44564]& 37938 [29419] & 53399 [52749] & 41543 [40867]\\
HT(r) &58862 [54324]& 45587 [34722] & 66936 [65883] & 51689 [51020]\\
HT(2r) & 78181 [75281]& 88327 [85689] & 106709 [106957] & 103036 [96124]\\

\textit{\underline{Pooled Curves}}\\
Local Poly & 40648 [32707]& 37345 [28707] & 41186 [40240] & 38084 [37464]\\
NWK& 41260 [33561]& 36791 [27848] & 41277 [40470] & 38092 [37468]\\
Smoothing Splines & 38737 [30617]& 36865 [27928] & 38281 [37269] & 36341 [35685]\\
Ensemble Ave &  50815 [44564]& 37938 [29419] & 53399 [52749] & 41543 [40867]\\

\textit{\underline{Curve - by - Curve}}\\
Global Kernel & 83134 [80354]& 43568 [36505] & 48418 [47520] & 38552 [38017]\\
Local Kernel & 62759 [58781]& 39777 [31632] & 45305 [44259] & 38780 [38184]\\
B-splines Regression &71699 [68131]& 38657 [30346] & 39255 [38373] & 39874 [39140]\\
Fourier Regression & 66629 [62591]& 54688 [49219] & 64238 [64015] & 51861 [51601]\\

\hline
\end{tabular}
\label{Tab}
\end{center}
\end{table}

Tables \ref{Tab:BB_SinCos} and \ref{Tab} contain the estimated mean squared errors for all the competing estimates, together with the hard-thresholded estimates; we also included the least squares estimator as a basis of comparison. For brevity we only included the results relative to $f(t)$ equal to Signal 1; we have obtained very similar results for Signal 2, and do not report them here for space considerations. The SNR for these simulations was set to 4.25, and we lower it substantially in the next section.  Our results support the following.  \\

\noindent {\it Conclusions on the performance of the fit of the estimators.}\\

\noindent 1. Our estimates, as expected, are sensitive to the basis choice. Since in this case the signal is differentiable, the best approximation and estimation is obtained via the Fourier basis. This supports, once more, the need for complementing any estimation procedure based on basis approximations with a basis selection step. \\

\noindent 2. If $\sigma^* = 10$, that is when the variance of the process dominates 
the variance of the random errors, the threshold estimators based on the Fourier basis perform essentially the same as most of the competing estimators, though they are consistently slightly better. It should be noted though that the performance of the competing estimators depends crucially on the selection of the tuning parameters of the respective method; therefore improvements of these estimates may be possible, via a refined choice of the respective tuning parameters. This may become very involved computationally and difficult to analyze theoretically. In contrast, our computationally simple estimator
is fully data-driven: the threshold levels are estimated from the data and our tuning parameter, the basis, can also be selected via a simple cross-validation method with proven optimality properties. \\

\noindent 3.  If $\sigma^* = 1$ the difference between our estimator and the competing ones is more pronounced, especially for the BM and ARIMA(1,1) processes, suggesting that this type of estimation is more robust against the variability in the data. As an additional remark, our experiments  indicate that some of the estimators proposed in the literature may be outperformed by the simple least squares estimator based on all the data points, or even by the naive sample average, if the choice of their tuning parameters is not refined; for all our simulations we did choose these tuning parameters adaptively as explained above, but we did not attempt to improve upon the published guidelines on their selection. 

\subsection{Simulation results: confidence bands}  The literature on confidence bands for the
 mean in model (\ref{model}) is limited. One stategy is to propose an estimator of $f(t)$, establish its asymptotic distribution, and use this to construct confidence intervals. Results pertaining to the asymptotic distribution of estimators of $f(t)$ is model (\ref{model}) are also limited; one exception is Zhang and Chen (2007), who studied the asymptotic distribution of the mean estimator $\wh f_{AK}(t)$ obtained by  averaging kernel smoothed individual trajectories. \\

The coverage of the 
bands obtained via such a strategy will be  affected by the accuracy 
of the estimators of the asymptotic variance. To investigate this effect we consider three  bands of the form: 
 \[ \left\{ \wh{f}_{AK}(t_{j}) \pm \sqrt{ \frac{{V(t_{j})}}{{n}}}z\bigg(\dfrac{\alpha}{2m}\bigg),\ 1 \le j \le m \right\}. \]  

\noindent  {\em Band 1}:   $V(t)=\Gamma(t,t)$, the theoretical variance function of the process $Z(t)$. \\
\noindent {\em Band 2}:  $V(t) = \wh{V}_{1}(t)$, where $\wh V_1(t)$ is estimated using functional principal components analysis. That is, $\wh{V}_{1}(t)=\sum_{r=1}^{R}\wh{\lambda}_{r} \wh{\xi}^{2}_{r}(t)$, where the estimated eigenvalues  $\wh{\lambda}_{r}=\wh{\text{Var}}(\xi_{r})$ and the estimated  eigenfunctions $\wh{\xi}_{r}$ are computed as in  Ramsay and Silverman (2005), section 8.4.2, and using the \texttt{fda} package in \texttt{R}. For our simulations we took $R = 10$.\\ 
\noindent  {\em  Band 3}:   $V(t) = \wh{V}_{2}(t)=\{1/(n-1)\}\sum_{i=1}^{n}(\wh{X}_{i}(t)-\wh{f}(t))^{2}$.  \\

We compare these bands with the proposed bands constructed as in Theorem \ref{Thm5}, (1) and (2), with Fourier basis functions $\phi_{k}$  ,$1\le k\le m$. We also investigate bands  based on  the untruncated least squares estimate: $\hat{f}(t_j) \pm \sum_{k=1}^m \wh r_k|\phi_k(t_j )|,\ 1\le j\le m$. 
The signal-to-noise ratio was set to 1.5 and 2.2, respectively.  Tables \ref{ARbands}  and \ref{BBbands} summarize the results for the AR(1) and BB processes and Signal 1.  We obtained results similar in spirit for all our other cases, and do not include them here for brevity. The entries in these tables are the widths of the confidence bands followed, in parentheses, by their empirical coverage. Recall that we are interested in simultaneous (over all $t_j$'s) coverage, and so  
the empirical coverage is given by  the relative frequency over simulations of
\begin{eqnarray}\label{an}
 \prod_{j=1}^m \1  \{ |f(t_j) - \wh g(t_j)|\le B_j\} \end{eqnarray}
for the various estimators $\wh g$ with their corresponding width $B_j$ at $t_j$. For instance, $\wh g= \wh f_{(\wh r)}$ and
 $B_j= \sum_k \wh r_k |\phi_k(t_j)| \{|\wh \mu_k| > \wh r_k\}$ correspond to our first proposed band in Tables \ref{ARbands}  and \ref{BBbands}. For our simulations we took $\alpha = 0.05$, and we therefore expect our bands to have at least $95 \%$ coverage. The results presented in Tables  \ref{ARbands}  and \ref{BBbands} below  support the following.  \\

\begin{table}
\tiny
\begin{center}
\caption{Width (Coverage) for confidence bands. Scenario: Signal 1, AR(1), $m=2^{6}=64$, $t \in [0,1]$ }
\label{Table 1}
\begin{tabular}{l  l  c c  c  c}
  \hline\\
 & $S = 500$ sims &    

   \multicolumn{2}{c}{$\text{signal-to-noise} = 1.5$}
   &
   \multicolumn{2}{c}{$\text{signal-to-noise}= 2.2$}

   \\

&  & $n=75$  & $n=100$ & $n=40$ & $n=50$\\
   \hline\\
    $\sigma^{*}=10$& & & & & \\
  & \textit{\underline{Bands based on asymp. normality}}\\
&Band 1 & 0.34 (0.97)  & 0.29 (0.97) & 0.46 (0.98) & 0.41 (0.97)   \\
&Band 2 &  0.31 (0.87)  & 0.26 (0.87) & 0.42 (0.86) & 0.37 (0.86)    \\
&Band 3 &  0.32 (0.94) & 0.27 (0.94)  & 0.43 (0.94) & 0.39 (0.94)   \\
 & &&&&\\
&\textit{\underline{Proposed Bands}}\\
&$\wh{f}_{(\wh{r})}(t_j) \pm \sum_{k=1}^m \hat{r}_k|\phi_k(t_j )| \II_{\{|\hat{\mu}_{k}|>\hat{r}_{k} \}}$ &  0.44 (0.96) & 0.40 (1.00) & 0.65 (0.97) & 0.61 (0.99)  \\
&$\wh{f}_{(\wh{r})}(t_j) \pm 3 \sum_{k=1}^m \hat{r}_k|\phi_k(t_j )| \II_{\{|\hat{\mu}_{k}|>\hat{r}_{k} \}}$   & 1.31 (1.00)  & 1.20 (1.00)  & 1.94 (1.00) & 1.84 (1.00) \\
&$\wh{f}_{(\wh{r})}(t_j) \pm \sum_{k=1}^m \hat{r}_k|\phi_k(t_j )|$ &  1.89 (1.00)  & 1.64 (1.00) & 2.58 (1.00) & 2.32 (1.00)  \\
& & &&&\\

\hline\\

    $\sigma^{*}=1$  && & & & \\
  & \textit{\underline{Bands based on asymp. normality}}\\
&Band 1 & 0.34 (0.25)  & 0.29 (0.08) & 0.46 (0.35) & 0.41 (0.21)   \\
&Band 2 &  0.33 (0.22)  & 0.29 (0.08) & 0.45 (0.30)  & 0.41 (0.22)    \\
&Band 3 &  0.33 (0.25) & 0.28 (0.08)  & 0.45 (0.36) & 0.41 (0.26)    \\
 & &&&&\\
 &\textit{\underline{Proposed Bands}}\\
& $\wh{f}_{(\wh{r})}(t_j) \pm \sum_{k=1}^m \hat{r}_k|\phi_k(t_j )| \II_{\{|\hat{\mu}_{k}|>\hat{r}_{k} \}}$ &  0.56 (0.99) & 0.52 (1.00) & 0.84 (0.99) & 0.78 (1.00)   \\
& $\wh{f}_{(\wh{r})}(t_j) \pm 3 \sum_{k=1}^m \hat{r}_k|\phi_k(t_j )| \II_{\{|\hat{\mu}_{k}|>\hat{r}_{k} \}}$   & 1.68 (1.00)  & 1.55 (1.00)  &  2.53 (1.00)& 2.35 (1.00)  \\
& $\wh{f}_{(\wh{r})}(t_j) \pm \sum_{k=1}^m \hat{r}_k|\phi_k(t_j )|$ &  3.15 (1.00)  & 2.73 (1.00) & 4.29 (1.00)  & 3.85 (1.00)  \\
 & & &&&\\
\hline
\end{tabular}
\label{ARbands}
\end{center}
\end{table}


\begin{table}
\tiny
\begin{center}
\caption{Width (Coverage) for confidence bands. Scenario: Signal 1, BB, $m=2^{6}=64$, $t \in (0,1)$ }
\label{Table 2}
\begin{tabular}{l  l  c c  c  c}
  \hline\\
 & $S = 500$ sims &    

   \multicolumn{2}{c}{$\text{signal-to-noise} = 1.5$}
   &
   \multicolumn{2}{c}{$\text{signal-to-noise}= 2.2$}

   \\

&  & $n=125$  & $n=150$ & $n=75$ & $n=100$\\
   \hline\\
    $\sigma^{*}=10$& & & & & \\
  & \textit{\underline{Competing Bands based on asymp. normality}}\\
&Band 1 & 0.23 (0.95)  & 0.21 (0.92)  &0.30 (0.93)  & 0.26 (0.90)   \\
&Band 2 &  0.23 (0.91)  & 0.21 (0.87)  & 0.29 (0.88)  &  0.25 (0.83)   \\
&Band 3 &  0.23 (0.93) & 0.21 (0.90)   & 0.29 (0.90) & 0.26 (0.86) \\
 & &&&&\\
&\textit{\underline{Proposed Bands}}\\
&$\wh{f}_{(\wh{r})}(t_j) \pm \sum_{k=1}^m \hat{r}_k|\phi_k(t_j )| \II_{\{|\hat{\mu}_{k}|>\hat{r}_{k} \}}$ & 0.33 (0.95)  & 0.31 (0.98) & 0.50 (0.94)  & 0.46 (0.99)   \\
&$\wh{f}_{(\wh{r})}(t_j) \pm 3 \sum_{k=1}^m \hat{r}_k|\phi_k(t_j )| \II_{\{|\hat{\mu}_{k}|>\hat{r}_{k} \}}$   &  0.99 (0.99)& 0.94 (0.99) & 1.51 (0.99)  & 1.37 (1.00)   \\
&$\wh{f}_{(\wh{r})}(t_j) \pm \sum_{k=1}^m \hat{r}_k|\phi_k(t_j )|$ &  1.14 (1.00)  & 1.04 (1.00)&  1.46 (1.00) & 1.27 (1.00)  \\
& & &&&\\

\hline\\

    $\sigma^{*}=1$  & &  & & & \\
  & \textit{\underline{Competing Bands based on asymp. normality}}\\
&Band 1 & 0.23 (0.00)   &  0.21 (0.00)  & 0.30 (0.00)  &  0.26 (0.00) \\
&Band 2 & 0.25 (0.00)   & 0.23 (0.00)  & 0.32 (0.00)  & 0.28 (0.00) \\
&Band 3 & 0.24 (0.00) & 0.22 (0.00)  & 0.31 (0.01) & 0.27 (0.00)   \\
 & &&&&\\
 &\textit{\underline{Proposed Bands}}\\
& $\wh{f}_{(\wh{r})}(t_j) \pm \sum_{k=1}^m \hat{r}_k|\phi_k(t_j )| \II_{\{|\hat{\mu}_{k}|>\hat{r}_{k} \}}$ & 0.47 (0.99)  & 0.44 (1.00)  & 0.63 (0.99)  & 0.55 (1.00)   \\
& $\wh{f}_{(\wh{r})}(t_j) \pm 3 \sum_{k=1}^m \hat{r}_k|\phi_k(t_j )| \II_{\{|\hat{\mu}_{k}|>\hat{r}_{k} \}}$   & 1.41 (1.00) &  1.32 (1.00)  & 1.88 (1.00)& 1.66 (1.00)  \\
& $\wh{f}_{(\wh{r})}(t_j) \pm \sum_{k=1}^m \hat{r}_k|\phi_k(t_j )|$ &2.22 (1.00)  & 2.02 (1.00)  & 2.86 (1.00) & 2.48 (1.00)  \\
 & & &&&\\
\hline
\end{tabular}
\label{BBbands}
\end{center}
\end{table}


\begin{figure}[htpb]
  \begin{center}
    \subfigure[\small Band 3 ]{\label{ a}\includegraphics[scale=0.4]{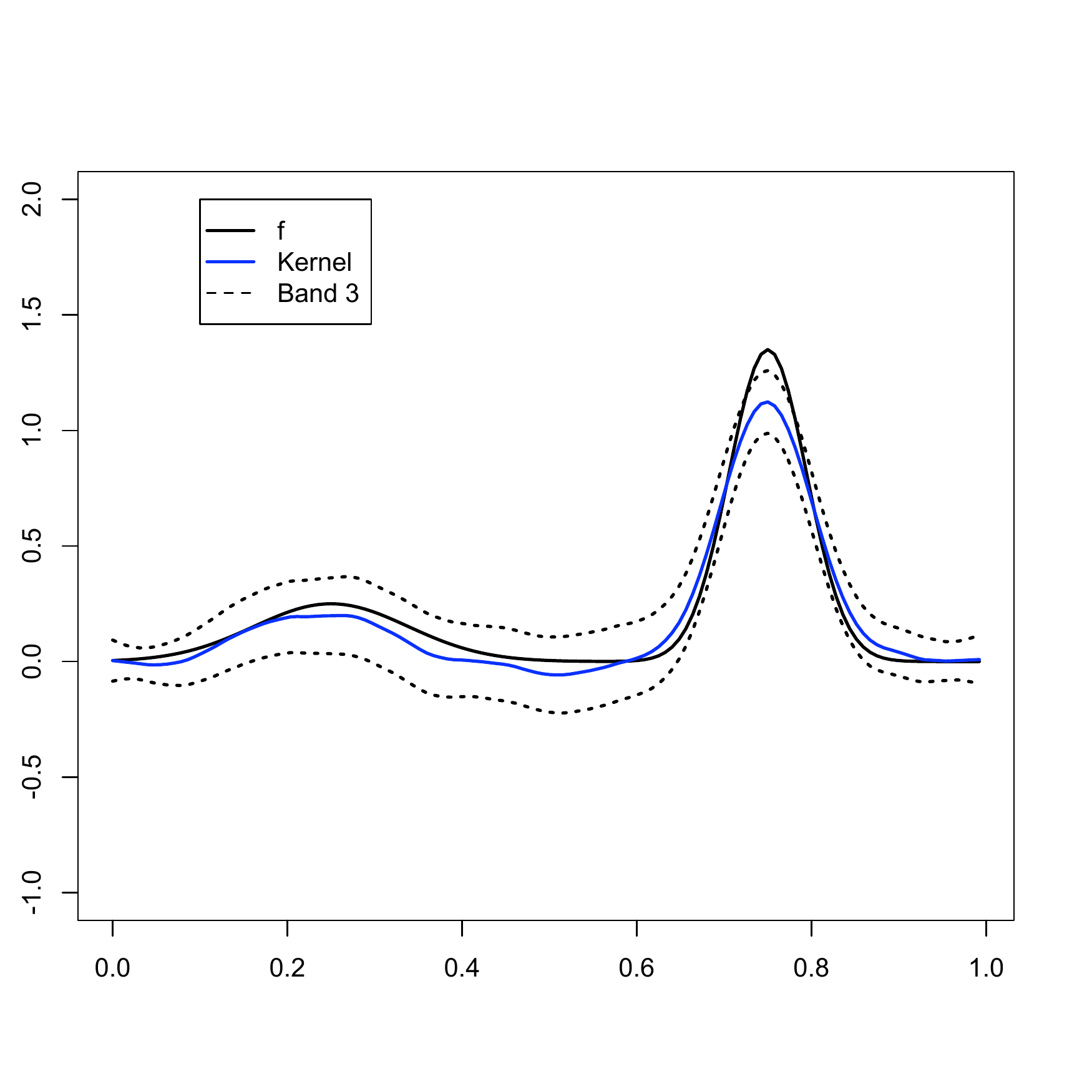}}
    \subfigure[\small Adaptive Band]{\label{b}\includegraphics[scale=0.4]{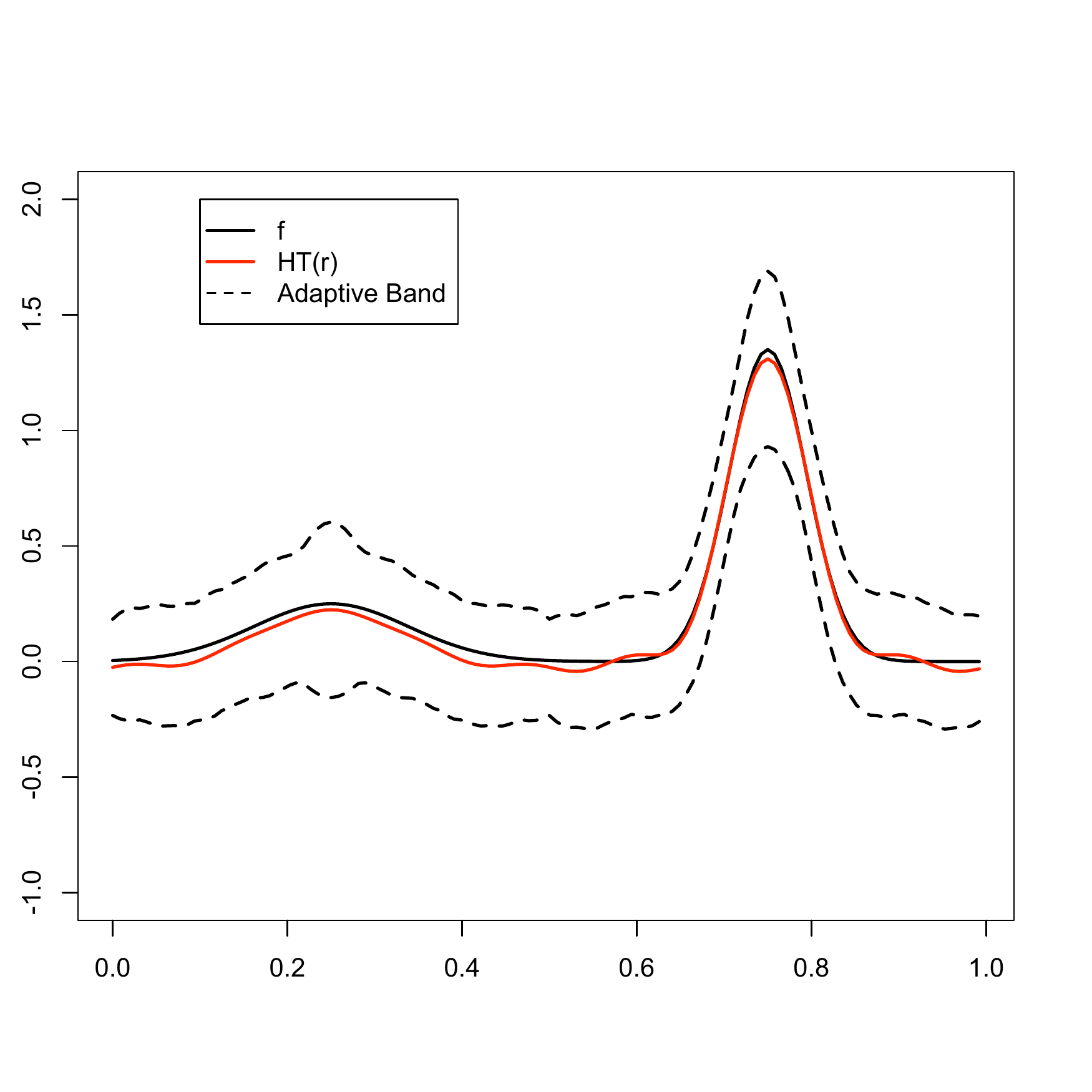}} 
    \subfigure[\small  Data + Adaptive Band]{\label{c}\includegraphics[scale=0.4]{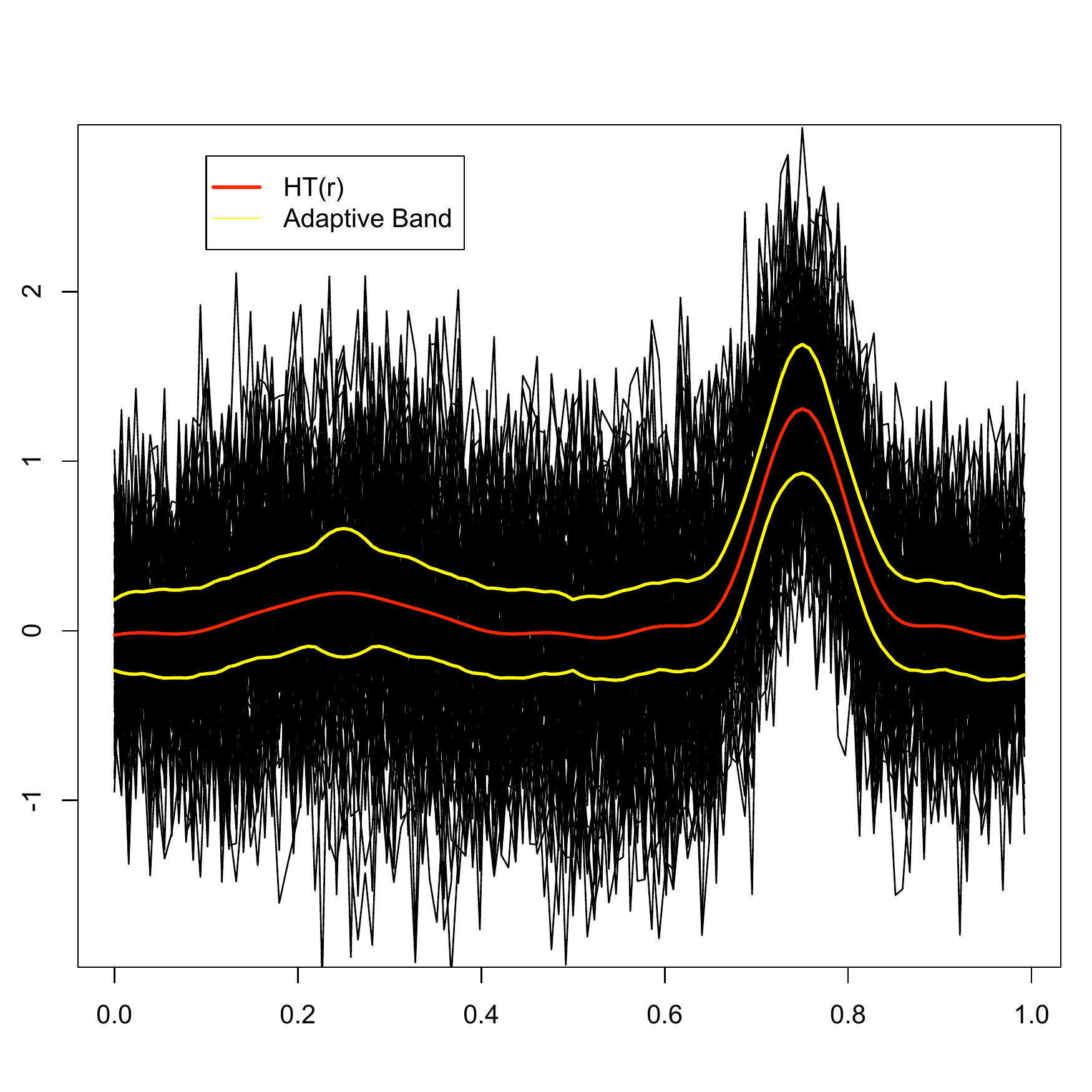}}\\
   \end{center}
  \caption{Scenario: BB, $\sigma^{*}=1$, SNR=2.2, $n=100$, $m=128$}
  \label{BB_1}
\end{figure}


\noindent {\it Conclusion on the proposed confidence bands } \\

\noindent 1. The coverage of the competing bands, based on the asymptotic normality of the estimators, depends crucially on the ratio of $\sigma_{\eps}^2$, the variance of the random errors $\varepsilon$, to the variance of the process, as measured by  $\sigma^*$ defined in (\ref{star}) above. This is expected, as the variance of the limiting distribution of these estimates depends only on  $\Gamma(t, t)$, and is independent of $\sigma_{\eps}^2$, which vanishes asymptotically in the limit.  Therefore,  bands that only take $\Gamma(t, t)$ into account cannot adapt to relatively large contributions of $\varepsilon$ to the overall variability, as quantified, in our simulations, by  $\sigma^* = 1$. \\
 
 Both Tables \ref{ARbands} and \ref{BBbands} show this phenomenon: the  bands based on asymptotic normality always become narrower as $n$ increases. However, since their width does not take into account $\sigma_{\eps}^2$, these bands loose completely (bottom of Tables \ref{ARbands} and  \ref{BBbands}) the excellent behavior they exhibit when $\sigma^* = 10$ (top of Tables \ref{ARbands} and \ref{BBbands}).  Figure \ref{BB_1} (a)  shows that for $\sigma^*=1$ the peak of $f$ is never covered by the band, which is the reason for the extremely poor uniform coverage; we recall that the coverage is computed via (\ref{an}).  In contrast, the first of our proposed bands is extremely robust to these variations, at the price of being slightly wider, as presented in  Figure \ref{BB_1} (b) - (c). \\

\noindent 2. Tables \ref{ARbands} and \ref{BBbands} support the expected fact that as the signal-to-noise ratio increases, our bands have good coverage for smaller sample sizes, as small as $n = 40$; since our bands are conservative we occasionally have $100 \%$ coverage.  The influence of $\sigma^*$ is, however, prevalent: even if the signal is stronger, 
the competing Bands 1 - 3 still cannot offer the required coverage 95 $\% $  for $\sigma^* = 1$, not even for the ideal band Band 1, that uses the theoretical variance of the processes. The differences between Band 2 and Band 3  illustrate the effect of estimating this variance on the coverage of the band.  \\

\noindent 3. The first of our proposed bands is uniform over  the space of parameters that are larger than the noise level quantified by $r_k$, $ 1 \leq k \leq m$. It is the least conservative of our bands, and is the one we suggest for use in practice. The trade-off between its width and its robustness against the variability in the data make it a strong competitor to the narrower, but less robust Bands 1 - 3. 
 The second of our bands is uniform over the whole parameter space and is therefore conservative and  necessarily wide; nevertheless it is always smaller, sometimes by almost a factor of two than the extreme case of a naive uniform band based only on the untruncated least squares estimator.

\bigskip

\appendix
\section{Proofs}

  \noindent Recall that we have introduced the following truncation levels which we repeat here for ease of reference. 
 \begin{eqnarray*}
 r_k = \frac{ z(\alpha/(2m)) }{\sqrt{n} } \sqrt{ \sigma_k^2 + \frac{\sigma_\eps^2}{m} }; \  \ \  \ \ \ \ 
 \wh r_k =  \frac{ z(\alpha/(2m)) }{\sqrt{n} } ( S_k + \delta)\\
 \bar  r_k = \frac{ z(\alpha/(2m)) }{\sqrt{n} } \left\{ \sqrt{ \sigma_k^2 + \frac{\sigma_\eps^2}{m} } + 2\delta \right\};  \ \ \ \ \ \ \
 \widetilde r_k = \frac{ z(\alpha/(2m)) }{\sqrt{n} } ( S_k +3\delta).
 \end{eqnarray*}
 \noindent Recall that the quantities $\wh \mu_k$ and $\mu_k$ are given respectively by (\ref{muhat}) and (\ref{mu}).

\begin{lemma}\label{lem}
Set \[ \Omega_{m,n,\delta} = \bigcap_{k=1}^m
 \left\{  |\wh \mu_k - \mu_k |\le r_k \right\} \cap \left\{ \wh r_k \ge r_k \right\} \cap \left\{ \wh r_k \le \bar r_k\right\}\cap \{ \bar r_k\le \widetilde r_k\}.
 \]
Then   \[ \lim_{\eps\downarrow0} \liminf_{n\to\infty}  \inf_{\mu_1,\ldots,\mu_m}\PP ( \Omega_{m,n,\delta}) \ge 1-\alpha\]
 for all $m$.\\
\end{lemma}

 \noindent {\it Remark.} Lemma \ref{lem} above is central to our analysis. The statements of Theorems \ref{Thm1} - \ref{Thm2} and \ref{Thm5} - \ref{Thm6} hold on the random set 
 $ \Omega_{m,n,\delta}$.  Lemma \ref{lem} shows that the probability of this set is larger than $1-\alpha$, asymptotically in $n$ and {\it uniformly} over $(\mu_1, \ldots, \mu_m)$, for any $m \geq 1$. \\
 
 \noindent {\bf Proof of Lemma \ref{lem}}. Set  $\bar A_k=(1/n)\sum_{i=1}^n A_{ik}$ and $\bar E_k=(1/n)\sum_{i=1}^n E_{ik}$ based on $A_{ik}$ defined above in (\ref{Aik}) and 
\[ E_{ik}= \frac1m \sum_{j=1}^m \eps_{ij} \phi_k(t_j),
\]
Then we can write for each fixed $m$, 
\begin{eqnarray*}
S_k^2 &=& \frac{1}{n-1} \sum_{i=1}^n ( \wh\mu_{i,k}- \wh\mu_k )^2\\ &=&
\frac{1}{n-1} \sum_{i=1}^n \left\{ (A_{ik} - \bar A_k) + (E_{ik}-\bar E_k)\right\}^2 .
\end{eqnarray*}
and consequently, $S_k^2\to \sigma_k^2 + \sigma_\eps^2/m$ almost surely, as $n\to\infty$. By the Bonferroni bound, the Central Limit Theorem, Kolmogorov's Strong Law of Large Numbers and Slutsky's Lemma, we have
 \[ \lim_{\eps\downarrow0} \liminf_{n\to\infty}  \inf_{\mu_1,\ldots,\mu_m}\PP ( \Omega_{m,n,\delta}) \ge 1-\alpha\]
 for all $m$, which is the desired result. $\blacksquare$\\
 
\subsection{Proof of Theorem \ref{Thm1}.} 
We first observe that
\begin{eqnarray*}
\|\wh f_{2r}-\bar f_r\|_{m,\infty} &=& \max_{1\le j\le m} \left| \sum_{k=1}^m (\wh\mu_k(2r_k) -\mu_k(r_k) )\phi_k(t_j) \right| \\
&\le& \max_{1\le k\le m} \|\phi_k\|_\infty   \sum_{k=1}^m  \left|\wh\mu_k(2r_k) -\mu_k(r_k)  \right| \end{eqnarray*}
and that 
\begin{eqnarray*}
\|\wh f_{2r}-\bar f_r\|_{m,2}^2 &=& \sum_{k=1}^m (\wh\mu_k(2r_k) -\mu_k(r_k) )^2 .
\end{eqnarray*}

It remains to show that on the event $\Omega_{m,n,\delta}$, \begin{eqnarray}\label{T1}
|\wh\mu_k(2 \wh r_k) -\mu_k(r_k)|\le 3 \bar r_k \1\{ |\mu_k|\ge r_k\}
\end{eqnarray} holds for all $1\le k\le m$, and any $m \geq 1$. Indeed, on $\Omega_{m,n,\eps}$,
\begin{eqnarray*}
&& \left| \wh\mu_k(2 \wh r_k) - \mu_k(r_k) \right| \\
&& = \left| \wh\mu_k\{ |\wh\mu_k| > 2 \wh r_k \} - \mu_k\{|\mu_k|>r_k\} \right| \\
&&\le |\wh\mu_k- \mu_k| \{ |\mu_k| > r_k \} +   |\wh\mu_k|  \left|\{ |\wh\mu_k| > 2\wh r_k \} - \{| \mu_k|>r_k\} \right| \\
&&\le   r_k\{ |\mu_k|>r_k\} + |\wh\mu_k| \left|\{ |\wh\mu_k|
> 2\wh r_k \} - \{| \mu_k|>r_k\} \right|
\end{eqnarray*}
For the second term, we consider two cases: $|\mu_k|\le r_k$ and
$|\mu_k|> r_k$ and we get
\begin{eqnarray*}
&& | \wh\mu_k| \left| \{ |\wh\mu_k| > 2\wh r_k \} - \{| \mu_k|>r_k\} \right|   \\
&&=   |\wh\mu_k| \{ |\wh\mu_k| >2\wh  r_k \}  \{ |\mu_k| \le r_k\} +
|\wh\mu_k| \{ |\wh\mu_k| \le 2\wh r _k \} \{|\mu_k|>r_k \}   \\
&&\le   2\wh r_k \{ |\mu_k|>r_k\}\\ &&\le   2\bar r_k \{ |\mu_k|>r_k\}
\end{eqnarray*}
where we used in the penultimate inequality the fact that $|\mu_k|\le r_k$ implies that $|\wh\mu_k|\le 2\wh r_k $. Combining the two preceding
displays yields (\ref{T1}). $\blacksquare$

\medskip

\subsection{Proof of Theorem \ref{Thm2}.}
 It suffices to show
that on the event $\Omega_{m,n,\delta}$, 
\begin{eqnarray}\label{T3}
|\widetilde\mu_k(2\wh r_k) -\mu_k(r_k)|\le 3\bar r_k \1\{ |\mu_k|\ge r_k\}
\end{eqnarray} holds for all $1\le k\le m$. The remainder of the proof is identical  to that  of Theorem \ref{Thm1}. To show (\ref{T3}), we begin by observing that
\begin{eqnarray*}
&& \left| \text{sgn}(\wh\mu_k)(|\wh\mu_k|-2\wh r_k)_{+} -\mu_k\{ |\mu_k|>r_k\}
\right| \\
&\le& |\text{sgn}(\wh\mu_k)-\text{sgn}(\mu_k)| |\mu_k| \{ |\mu_k|> r_k\} +
\left| (|\wh\mu_k|-2\wh r_k)_{+} -|\mu_k|\{ |\mu_k|>r_k\} \right| \end{eqnarray*}
The first term on the right is zero:
\begin{eqnarray*}
&& |\text{sgn}(\wh\mu_k)-\text{sgn}(\mu_k)| |\mu_k| \{ |\mu_k|> r_k\}\\
&&=\{\mu_k> r_k\} |\text{sgn}(\wh\mu_k)-1| | \mu_k|  + \{ \mu_k<-r_k\} |\text{sgn}(\wh\mu_k)+1| |\mu_k| \\
&&=0
\end{eqnarray*} since
 $\mu_k>r_k$ (and $ |\mu_k-\wh\mu_k| \le r_k$) implies that $\wh\mu_k>0$ and
sgn$(\wh\mu_k)=1$, and, in a similar way, $\mu_k<-r_k$ implies that $\wh\mu_k<0$ and
sgn$(\wh\mu_k)= -1$. Consequently,
\begin{eqnarray*}
  \left|\text{sgn}(\wh\mu_k)(|\wh\mu_k|-2\wh r_k)_{+} - \mu_k\{ |\mu_k|>r_k\} \right| 
&\le& \left| (|\wh\mu_k|-2\wh r_k)_{+} -|\mu_k|\{ |\mu_k|>r_k\} \right| \end{eqnarray*}
Next, since $|\mu_k|\le r_k$ implies that $|\wh\mu_k|\le 2r_k\le 2\wh r_k\le 2\bar r_k$, we obtain 
\begin{eqnarray*}
&& \left| (|\wh\mu_k|-2\wh r_k)_+ - |\mu_k|\{ |\mu_k|>r_k\} \right|\\
&&\le |\mu_k| \{ |\mu_k|>r_k,\ |\wh\mu_k|\le 2\wh r_k\} +
\left| |\wh\mu_k|-|\mu_k|-2\wh r_k \right| \{ |\mu_k|>r_k,\ |\wh\mu_k|> 2\wh r_k\}\\
&&\le 3 \bar r_k \{ |\mu_k|>r_k\}.
\end{eqnarray*} 
Combination of all these bounds gives (\ref{T3}). $\blacksquare$

\medskip

\subsection{Proof of Theorem \ref{Thm5}.}
We first notice that on the event $\Omega_{m,n,\delta}$, we
have 
\begin{eqnarray}\label{T2}
|\wh\mu_k(\wh r_k) -\mu_k(2\bar r_k)|\le 3 \widetilde r_k \1\{ |\wh\mu_k|\ge\wh r_k\}
\end{eqnarray} for all $1\le k\le m$. 
This follows essentially from interchanging the roles of  $\wh\mu_k$ and $\mu_k$ and $\bar r_k$ and $\wh r_k$ in the proof of (\ref{T1}):
\begin{eqnarray*}
&& \left| \mu_k(2\bar r_k) - \wh \mu_k(\wh r_k) \right| \\
&& = \left| \mu_k\{ | \mu_k| > 2\bar r_k \} - \wh \mu_k\{|\wh \mu_k|>\wh r_k\} \right| \\
&&\le |\wh\mu_k- \mu_k| \{ |\wh \mu_k| > \wh  r_k \} +   |\mu_k|  \left|\{ |\mu_k| > 2\bar r_k \} - \{|\wh \mu_k|>\wh r_k\} \right| \\
&&\le   r_k\{ |\wh\mu_k|>\wh r_k\} + |\mu_k| \left|\{ |\mu_k|
> 2\bar r_k \} - \{| \wh\mu_k|>\wh r_k\} \right|
\\
&&  r_k\{ |\wh\mu_k|>\wh r_k\} +| \mu_k| \left| \{ | \mu_k| > 2 \bar r_k \} - \{|\wh \mu_k|>\wh r_k\} \right|   \\
&&=  r_k\{ |\wh\mu_k|>\wh r_k\} +  |\mu_k| \{ |\mu_k| >2\bar r_k \}  \{ |\wh \mu_k| \le\wh r_k\} +
|\mu_k| \{ |\mu_k| \le 2\bar r _k \} \{|\wh \mu_k|>\wh r_k \}   \\
&&\le  r_k\{ |\wh\mu_k|>\wh r_k\} +  2\bar r_k \{ |\wh \mu_k|>\wh r_k\}\\
&&\le 3\widetilde r_k \{ |\wh \mu_k|>\wh r_k\}
\end{eqnarray*}
Since
\begin{eqnarray*}
&&\PP\left\{ |\wh f_{(\wh r)}(t_j) - \bar f_{(2\bar r)}(t_j)| \le 3 \sum_{k=1}^m \widetilde r_k |\phi_k(t_j) \1\{ |\wh\mu_k|\ge\wh r_k\} ,\ 1\le j\le m\right\}\\
&&\ge\PP\left\{ |\wh\mu_k(\wh r_k) -\mu_k(2\bar r_k)|\le 3\widetilde r_k \1\{ |\wh\mu_k|\ge\wh r_k\},\ 1\le k\le m\right\}\\
&&\ge \PP(\Omega_{m,n,\delta}).
\end{eqnarray*}
The second claim follows from
\begin{eqnarray*}
|\wh\mu_k(\wh r_k) -\mu_k(2\bar r_k)| &\le& r_k \1\{ |\wh\mu_k|>\wh r_k\} + |\mu_k|1\{  |\mu_k|\le 2\bar r_k\}\1\{|\wh\mu_k|>\wh r_k\}\\
&=& r_k\1\{ |\wh\mu_k| >\wh r_k\}
\end{eqnarray*}
if we have that all $\mu_k = 0$ if $|\mu_k|\le 2\bar r_k$. $\blacksquare$ \\

\subsection{Proof of Theorem \ref{Thm6}.}
The proof is similar to the one of Theorem \ref{Thm5}, except that we replace (\ref{T2}) by the following inequality, which holds on the event $\Omega_{m,n,\delta}$:
\begin{eqnarray*}
&&\left| \text{sign}(\widehat \mu_k)\left(|\wh\mu_k|-\wh  r_k \right)_{+}  - \text{sign}(\mu_k) |\mu_k| \1\{ |\mu_k|\ge 2\bar r_k\} \right| \\
 &&   \hspace{0.3cm} \le  \left| \text{sign}(\widehat \mu_k) -\text{sign}(\mu_k) \right| |\mu_k|\1\{|\mu_k|\ge 2\bar r_k\}+
\left|
\left(|\wh\mu_k|- \wh r_k \right)_{+}  - |\mu_k| \1\{ |\mu_k|\ge 2\bar r_k\} \right| \\
&&\hspace{0.3cm}=\left|
\left(|\wh\mu_k|- \wh r_k \right)_{+}  - |\mu_k| \1\{ |\mu_k|\ge 2\bar r_k\} \right|\\
&&\hspace{0.3cm}\le   \left|
\left(|\wh\mu_k|- \wh r_k \right)_{+}  - |\mu_k|  \right|  \1\{ |\mu_k|\ge 2\bar r_k\}+  
\left(|\wh\mu_k|-\wh r_k \right)_{+}  \1\{ |\mu_k|< 2\bar r_k\} \\ 
&&\hspace{0.3cm}=\left| \left(|\wh \mu_k|-\wh r_k\right)_+ -|\mu_k| \right| \1\{|\mu_k|\ge 2\bar r_k\}\1\{|\wh\mu_k|>\wh r_k\}  \\
&&\hspace{0.5cm} + \left(|\wh \mu_k|-\wh r_k\right)_+ \1\{| \wh\mu_k| \le 3\widetilde r_k\}\1\{ |\mu_k|<2\bar r_k\}\\
&&\hspace{0.3cm}\le 2\wh r_k \1\{|\mu_k|\ge 2\bar r_k\}\1\{|\wh\mu_k|>\wh r_k\} + 2\widetilde r_k \1\{\wh  r_k<| \wh\mu_k| \le3\widetilde  r_k\}\1\{ |\mu_k|<2\bar r_k\}\\
&& \hspace{0.3cm}\le 2\widetilde r_k\{ |\wh\mu_k|>\wh r_k\}. \ \ \ \blacksquare
\end{eqnarray*} 
\subsection{Proof of Theorem \ref{Thm3}}
First observe that
\begin{eqnarray*}
&& \EE \| \wh f_{(2r)}- \bar f_{(r)} \|_{m,2}^2 = \sum_{k=1}^m \EE |\wh \mu_k(2r_k)-\mu_k(r_k)|^2\\
&& =  \sum_{k=1}^m \EE \left[ (\wh \mu_k-\mu_k) I\{ |\mu_k|> r_k\} + \wh\mu_k(I\{ |\wh\mu_k|> 2r_k\} - I\{ |\mu_k|> r_k\} )\right]^2\\
&& \le 2\sum_{k=1}^m  (\frac{\sigma_k^2}{n}+\frac{\sigma_\eps^2}{mn}) I\{ |\mu_k|>r_k\} + \sum_{k=1}^m 2 \EE \wh \mu_k^2 I\{|\wh \mu_k|>2r_k,\ |\mu_k|\le r_k\}\\
&& \ \ + \sum_{k=1}^m 2 \EE \wh \mu_k^2 I\{|\wh \mu_k|\le 2r_k,\ |\mu_k|> r_k\} 
\\
&& \le 2 \sum_{k=1}^m  (\frac{\sigma_k^2}{n}+\frac{\sigma_\eps^2}{mn}) I\{ |\mu_k|>r_k\} \\
&&+ \sum_{k=1}^m 4\EE  \left[ \mu_k^2 + (\wh \mu_k-\mu_k)^2 \right]  I\{|\wh \mu_k|>2r_k,\ |\mu_k|\le r_k\}\\
&&+ \sum_{k=1}^m 2(2r_k)^2 I\{|\wh \mu_k|\le 2r_k,\ |\mu_k|> r_k\} .
\end{eqnarray*}
Using the triangle inequality readily yields the first claim.
For the second claim, it is assumed that $(\wh \mu_k-\mu_k) / \sqrt{\sigma_k^2+\sigma_\eps^2/m} $ is $N(0,1)$.
Hence
\begin{eqnarray*}
 \EE \left[ (\wh \mu_k-\mu_k)^2 I\{ |\wh \mu_k -\mu_k|> r_k\} \right]
&=& \left(\frac{ \sigma_k^2}{n} + \frac{\sigma_\eps^2}{n m} \right ) \int _{|x|>z(\alpha/(2m))} x^2 \varphi(x)\, dx\\
&=&   \left(\frac{ \sigma_k^2}{n} + \frac{\sigma_\eps^2}{n m} \right ) \int_{|x|>z(\alpha/(2m))} \varphi(x)\, dx\\
&=&   \left(\frac{ \sigma_k^2}{n} + \frac{\sigma_\eps^2}{n m} \right )
 \frac{\alpha}{m}
\end{eqnarray*}
so that
\begin{eqnarray*}
 4 \sum_{k=1}^m \EE\left[  \left( r_k^2 +    (\wh \mu_k-\mu_k) ^2  \right) I\{|\wh \mu_k-\mu_k|> r_k\}\right] \\
= \frac{4\alpha}{m} \sum_{k=1}^m\left ( r_k^2+\frac{\sigma_k^2}{n}+\frac{\sigma_\eps^2}{nm}\right )
\end{eqnarray*}
as claimed. $\blacksquare$

\medskip
\subsection{Proof of Theorem \ref{ds}}
 For any real function $g$, that is independent of $Y_{ij}$, $i\in I_2$, we have
 \begin{eqnarray*}
 S_2( g) = \EE \left[ \wh S_2( g) \right] = \| f-\wh g\|_{m,2}^2 +\frac1m \sum_{j=1}^m \Gamma(T_j,T_j) + \sigma_\eps^2
 \end{eqnarray*}
 so that
 \[ S_2(g)- S_2(f) = \| g-f\|_{m,2}^2.\]
 Repeating the same arguments as in Lemma 2.3 of Wegkamp (2003), we obtain
 \begin{eqnarray*}
 \| \wh g- f\|_{m,2}^2 &\le& (1+a) \| \wh g_\ell - f\|_{m,2}^2 \\ &&+\max_{\ell} \left[
 2(1+a)\frac{1}{n_2} \sum_{i\in I_2} \frac1m \sum_{j=1}^m \{ \eps_{ij} + Z_i(T_j)
\} \{ \wh g_\ell (T_j)- f(T_j)\} - a\|\wh g_\ell -f\|_{m,2}^2 \right]
\end{eqnarray*} 
for all $a>0$ and all $k\ge1$.
Furthermore, using the inequality $2xy \le x^2/c + cy^2$ for
$c=(1+a)/a$, $y=\| \wh g_\ell -f\|_{m,2}$, \[ x=\frac{1}{n_2} \sum_{i\in I_2} \frac1m \sum_{j=1}^m \{ \eps_{ij} + Z_i(T_j)
\} \frac{\{ \wh g_\ell (T_j)- f(T_j)\} }{\| \wh g_\ell -f\|_{m,2}}
\]
we obtain
\begin{eqnarray*}
 \| \wh g - f\|_{m,2}^2&\le& (1+a) \| \wh g_\ell - f\|_{m,2}^2 \\ &&+\max_{\ell} 
 \frac{(1+a)^2}{a} \left[ \frac{1}{n_2} \sum_{i\in I_2} \frac1m \sum_{j=1}^m \{ \eps_{ij} + Z_i(T_j)
\}  \frac{ \{ \wh g_\ell(T_j)- f(T_j)\} }{\|\wh g_\ell -f\|_{m,2}} \right]^2  \\
&\le& (1+a) \| \wh g_\ell - f\|_{m,2}^2 \\ &&+\max_{\ell} 
2 \frac{(1+a)^2}{a} \left[ \frac{1}{n_2} \sum_{i\in I_2} \frac1m \sum_{j=1}^m  Z_i(T_j)
  \frac{ \{ \wh g_\ell(T_j)- f(T_j)\} }{\|\wh g_\ell -f\|_{m,2}} \right]^2 \\
&&+\max_{k} 
2 \frac{(1+a)^2}{a} \left[ \frac{1}{n_2} \sum_{i\in I_2} \frac1m \sum_{j=1}^m \eps_{ij} 
  \frac{ \{ \wh g_\ell (T_j)- f(T_j)\} }{\|\wh g_\ell -f\|_{m,2}} \right]^2
\end{eqnarray*}
By Rosenthal's (see, e.g., Wegkamp 2003, page 262), the  Cauchy-Schwarz and Jensen's inequalities, we obtain
\begin{eqnarray*}
&& \EE  \left[ \frac{1}{n_2} \sum_{i\in I_2} \frac1m \sum_{j=1}^m \eps_{ij} 
  \frac{ \{ \wh g_\ell (T_j)- f(T_j)\} }{\|\wh g_\ell -f\|_{m,2}} \right]^p\\
  &&\le C_p n_2^{-p} \max\left(   \sum_{i\in I_2}\EE \left| \frac1m \sum_{j=1}^m \eps_{ij} 
  \frac{ \{ \wh g_\ell(T_j)- f(T_j)\} }{\|\wh g_\ell -f\|_{m,2}} \right|^p ,  \left[ \sum_{i\in I_2} \EE \left| \frac1m \sum_{j=1}^m \eps_{ij} 
  \frac{ \{ \wh g_\ell(T_j)- f(T_j)\} }{\|\wh g_\ell -f\|_{m,2}} \right|^2 \right]^{p/2}
\right)\\
&&\le C_p n_2^{-p} \max\left(   \sum_{i\in I_2}\EE \left| \frac1m \sum_{j=1}^m \eps_{ij} 
 ^2 \right|^{p/2}  ,  \left[ \sum_{i\in I_2} \EE \frac1m \sum_{j=1}^m \eps_{ij} 
 ^2 \right]^{p/2}
\right)\\
&&\le C_p n_2^{-p} \max\left(   \sum_{i\in I_2}  \frac1m \sum_{j=1}^m\EE | \eps_{ij}|^p 
   ,  \left[ \sum_{i\in I_2} \EE \frac1m \sum_{j=1}^m \eps_{ij} 
 ^2 \right]^{p/2}
\right)\\
&&= C_p n_2^{-p} \max(n_2  \tau_{\eps,p}, (n_2 \bar \sigma_\eps^2)^{p/2} ).
\end{eqnarray*}
In the same way,
\begin{eqnarray*}
&& \EE  \left[ \frac{1}{n_2} \sum_{i\in I_2} \frac1m \sum_{j=1}^m Z_i(T_j)
  \frac{ \{ \wh g_\ell (T_j)- f(T_j)\} }{\|\wh g_\ell -f\|_{m,2}} \right]^p\\
&&= C_p n_2^{-p} \max(n_2 \bar \tau_{Z,p}, (n_2 \bar \tau_{Z,2})^{p/2} ).
\end{eqnarray*}
Using the same arguments as in the proofs of Lemma 2.5 and Theorem 2.2 in Wegkamp (2003), we arrive at
\begin{eqnarray*}
 &&\max_{k} 
 \frac{(1+a)^2}{a} \left[ \frac{1}{n_2} \sum_{i\in I_2} \frac1m \sum_{j=1}^m \{ \eps_{ij} + Z_i(T_j)
\}  \frac{ \{ \wh g_\ell (T_j)- f(T_j)\} }{\|\wh g_\ell -f\|_{m,2}} \right]^2  \\
&\le&\frac{1+a}{n} +(1+a) K \frac{ C_p}{n} \left(\bar\tau_{\eps,p}+\bar\sigma_\eps^p+ \bar\tau_{Z,p}+\bar\tau_{Z,2}^{p/2} \right) \left( \frac{1+a}{a}\right)^{p/2} 
\end{eqnarray*}
and take $a=1$ to obtain the result.
\qed\\

\end{document}